\documentclass[%
reprint,amsmath,amssymb,aps,superscriptaddress
]{revtex4-2}
\usepackage{graphicx}  
\usepackage{dcolumn}   
\usepackage{bm}        
\usepackage[colorlinks]{hyperref}
\usepackage[all]{hypcap}
\usepackage{color}
\usepackage{amsmath}
\usepackage{physics}

\newcommand{\red}[1]{{\color{black}#1}}

\newcommand{\cqt}{Centre for Quantum Technologies, National University of Singapore, 3 Science Drive 2, Singapore 117543, Singapore}
\newcommand{\ntu}{Division of Physics and Applied Physics, School of Physical and Mathematical Sciences, Nanyang Technological University, 21 Nanyang Link, Singapore 637371, Singapore}

\begin{document}
\preprint{APS/123-QED}
\title{ICARUS-Q: Integrated Control and Readout Unit for Scalable Quantum Processors}

\author{Kun Hee \surname{Park}} 
\affiliation{\cqt{}}

\author{Yung Szen \surname{Yap}}
\email[The author to whom correspondence may be addressed: ]{yungszen@utm.my}
\affiliation{Faculty of Science and Centre for Sustainable Nanomaterials (CSNano), Universiti Teknologi Malaysia, 81310 UTM Johor Bahru, Johor, Malaysia}
\affiliation{\cqt{}}

\author{Yuanzheng Paul \surname{Tan}}
\affiliation{\ntu{}}
\affiliation{\cqt{}}

\author{Christoph \surname{Hufnagel}}
\affiliation{\cqt{}}

\author{Long Hoang \surname{Nguyen}}
\affiliation{\ntu{}}

\author{Karn Hwa \surname{Lau}}
\affiliation{Advinno Technologies Pte.~Ltd., 22, Sin Ming Lane, \#05-75 Midview City
573969, Singapore}

\author{Patrick \surname{Bore}}
\affiliation{\cqt{}}

\author{Stavros \surname{Efthymiou}}
\affiliation{Quantum Research Centre, Technology Innovation Institute, Accelerator 2 building, on Plot M12, PO Box 9639, Masdar City, Abu Dhabi, UAE}

\author{Stefano \surname{Carrazza}}
\affiliation{TIF Lab, Dipartimento di Fisica, Universit\`a degli Studi di Milano and INFN Sezione di Milano, via Celoria 16, 20133, Milan, Italy}
\affiliation{Theoretical Physics Department, CERN, Esplanade des Particules 1, 1211 Meyrin, Switzerland}
\affiliation{Quantum Research Centre, Technology Innovation Institute, Accelerator 2 building, on Plot M12, PO Box 9639, Masdar City, Abu Dhabi, UAE}

\author{Rangga P. \surname{Budoyo}}
\affiliation{\cqt{}}

\author{Rainer \surname{Dumke}}
\email{rdumke@ntu.edu.sg}
\affiliation{\cqt{}}
\affiliation{\ntu{}}

\begin{abstract}
We present a control and measurement setup for superconducting qubits based on Xilinx 16-channel radio-frequency system-on-chip (RFSoC) device.  
The proposed setup consists of four parts: multiple RFSoC boards, a setup to synchronise every digital to analog converter (DAC), and analog to digital converter (ADC) channel across multiple boards, a low-noise direct current (DC) supply for tuning the qubit frequency and cloud access for remotely performing experiments. 
We also design the setup to be free of physical mixers. The RFSoC boards directly generate microwave pulses using sixteen DAC channels up to the third Nyquist zone which are directly sampled by its eight ADC channels between the fifth and the ninth zones.

\end{abstract}
\maketitle


\section{Introduction}

Superconducting qubits in the dilution refrigerator are controlled and measured with room temperature electronics. A typical superconducting qubit is designed with its transition energy in the order of a few GHz, and requires arbitrary and precise microwave generation and detection for control and measurement. As the number of qubits increases, the number of microwave channels required increases linearly. Therefore, designing a qubit control system that is scalable, compact and cost-effective, while maintaining its precision, speed, and features, is imperative. 

Apart from the microwave circuits for frequency up/down-conversion, a basic qubit control system consists of digital to analog converters (DACs), and analog to digital converters (ADCs) and stable current sources; the DACs generate the microwave pulses that travel into the fridge, the ADCs digitise the analog signals that travel out from the fridge, and the current source tunes the qubit frequencies. Some of the earlier microwave control systems for electron spin and superconducting qubits~\cite{esr}, relied on benchtop arbitrary waveform generators (AWGs) for the microwave generation~\cite{tseitlin2011digital, doll2013adiabatic, ku2015band, riste2015detecting, raftery2017direct, zaw2021ghost}. The recent trends however are favoring field programmable gate array (FPGA)~\cite{kaufmann2013dac, ryan2017hardware, salathe2018low, lin2018high,sun2020scalable, kalfus2020high, xu2021qubic, yang2021fpga} for their higher number of channels (i.e.~cost per channel), versatility, and the form factor. Typically, two DAC channels are required, per qubit, for qubit driving; additionally, one DAC channel is shared among five or more qubits for frequency-multiplexed readout schemes ~\cite{lanting2005frequency, chen2012multiplexed, hornibrook2014frequency}. 

The latest family of FPGA by Xilinx known as the Zynq UltraScale+ radio-frequency system-on-chip (RFSoC)~\cite{xilinxWebsite} hosts a wide-variety of features that are advantages for qubit control and measurement.  To the best of our knowledge, this family of devices feature the highest number of independent DAC and ADC channels within a single chip with high sampling rates and is equipped auto-synchronisation between channels. The device also has digital up/down converters using internal complex mixers and a 48-bit numerically-controlled oscillator (NCO), and two processors.
These features which are available at a fraction of the cost and size to that of other commercial off-the-shelf devices, makes the RFSoC particularly enticing for applications such as radar~\cite{phased2018radar},  communications~\cite{farley2017programmable} and quantum computing~\cite{gebauer2020state, stefanazzi2021qick, esposito2021observation}. 

First announced in late 2018, the RFSoC has three generations of devices where only the first two are available at the time of writing~\cite{secondgen}. Within the first generation of RFSoC devices, the two top devices are XCZU28DR and XCZU29DR.  There primary differences between them are: the number of channels (8 versus 16~channels of DACs and ADCs) and the maximum sampling rate of the ADCs (4.096~GS/s versus 2.058~GS/s, respectively). 

A single RFSoC board (ZCU111 by Xillinx) that is populated with the XCZU28DR (equipped with eight 6.554~GS/s DAC channels and eight 4.096~GS/s ADC channels) has been used as a control and measurement system for superconducting quantum computers~\cite{gebauer2020state, stefanazzi2021qick, esposito2021observation}. Here, we develop a scalable setup based on multiple synchronised XCZU29DR RFSoC boards, where each board features sixteen 6.554~GS/s DAC channels and sixteen 2.058~GS/s ADC channels per board and operates without physical IQ mixers.

\section{Implementation}
\label{sec:implementation}
\begin{figure*}[!ht]
	\centering
	\includegraphics[width=\linewidth]{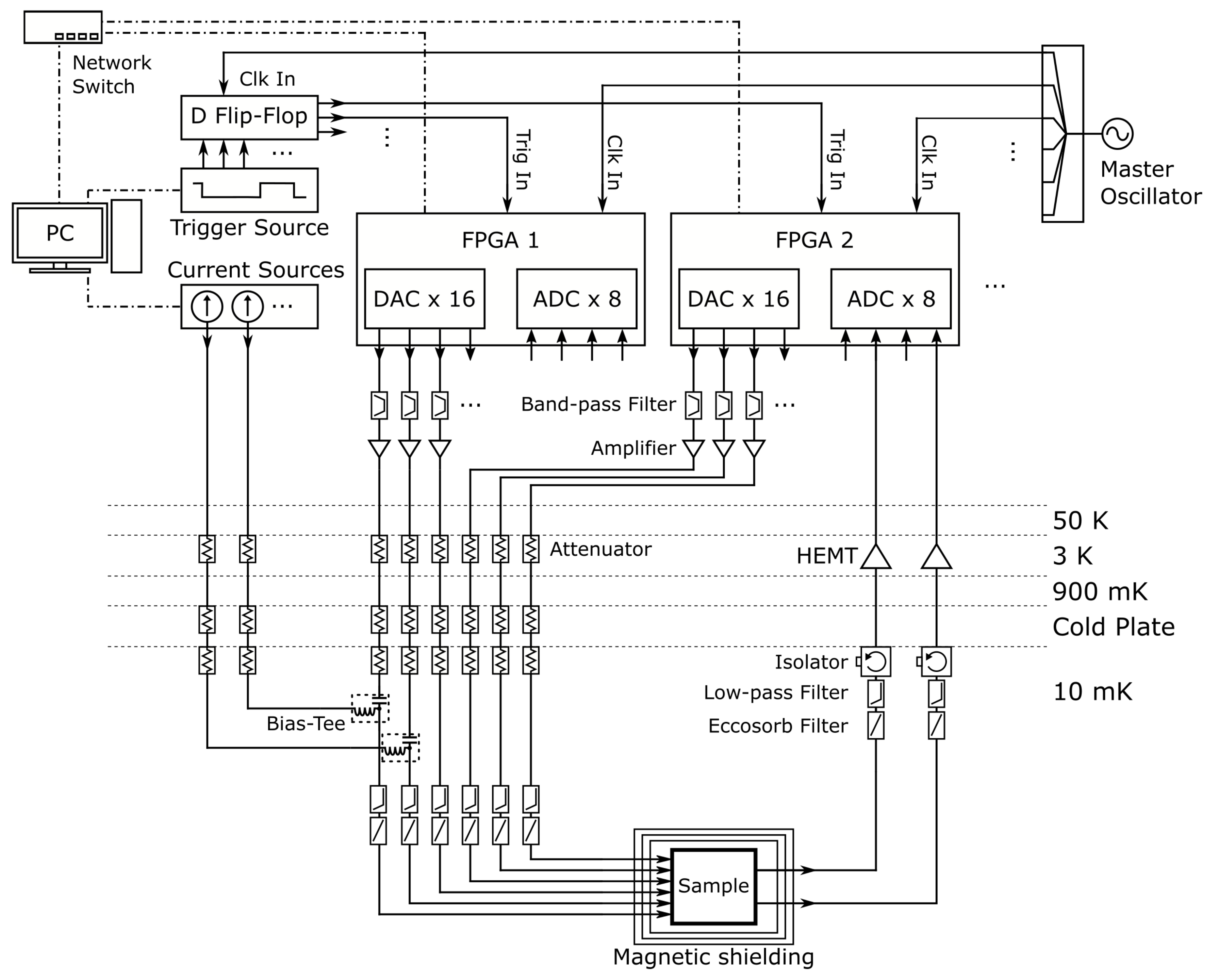}
	\caption{Overall circuit diagram. Several units of RFSoC boards, each with an RFSoC device (XCZU29DR by Xilinx), are used to directly generate and sample microwave signals in the GHz frequencies using its higher Nyquist zones (with the help of filters). The boards are synchronised to a master reference clock (VHF Citrine Gold, Wenzel) and triggers from a programmable~\cite{PBpython} TTL pulse generator (PulseBlaster PB24-100-4K, Spincore). The triggers are also synchronised to the same clock using a d-type flip-flop circuit. The trigger source board runs a custom firmware which takes in the master oscillator signal as the reference clock. To tune the superconducting qubits, low noise current sources are merge with microwave pulses using bias-tees, located at the mixing chamber stage. The biasing for the flip-flop clock input has been omitted for clarity. The worker PC is connected to the trigger source via PCIe, and to the current sources via USB. The RFSoC boards are controlled from the PC via SSH over Ethernet through the network switch.}
	\label{pic:overall}
\end{figure*}

The developed setup in Fig.~\ref{pic:overall} consists of several parts. 
The RFSoC board, codename ICARUS-Q (Integrated Control And Readout Unit for Scalable Quantum processors), runs with an embedded Linux kernel that receives commands and transfers data in/out of the board via Ethernet.
In our approach, an alias in the higher Nyquist zone~\cite{kalfus2020high} of the DAC signal is used to address the GHz-range qubit transition and direct sampling of the high-frequency signals by the ADCs at a lower sampling rate (Sec.~\ref{sec:FPGA} -- \ref{sec:feedback}). Multiple RFSoC boards are synchronised with each other using a master oscillator and trigger signals that are synchronised to the same master oscillator (Sec.~\ref{sec:sync}). 
For tunable superconducting qubits, the Josephson junction of the qubit is replaced by a DC-SQUID loop. This allows the qubit to be tuned with a magnetic flux, which is coupled into the loop from a current carrying wire close to the loop.
To support this, low-noise DC sources are integrated into the setup for tuning the qubits (Sec.~\ref{sec:dcbias}). 
The RFSoC boards, trigger and current sources are connected to a PC, which runs a worker program that communicates with a cloud server to remotely run experiments ({Sec.~\ref{sec:cloud}}).

\begin{figure}[!h]
	\centering
    \includegraphics[height=4.5cm]{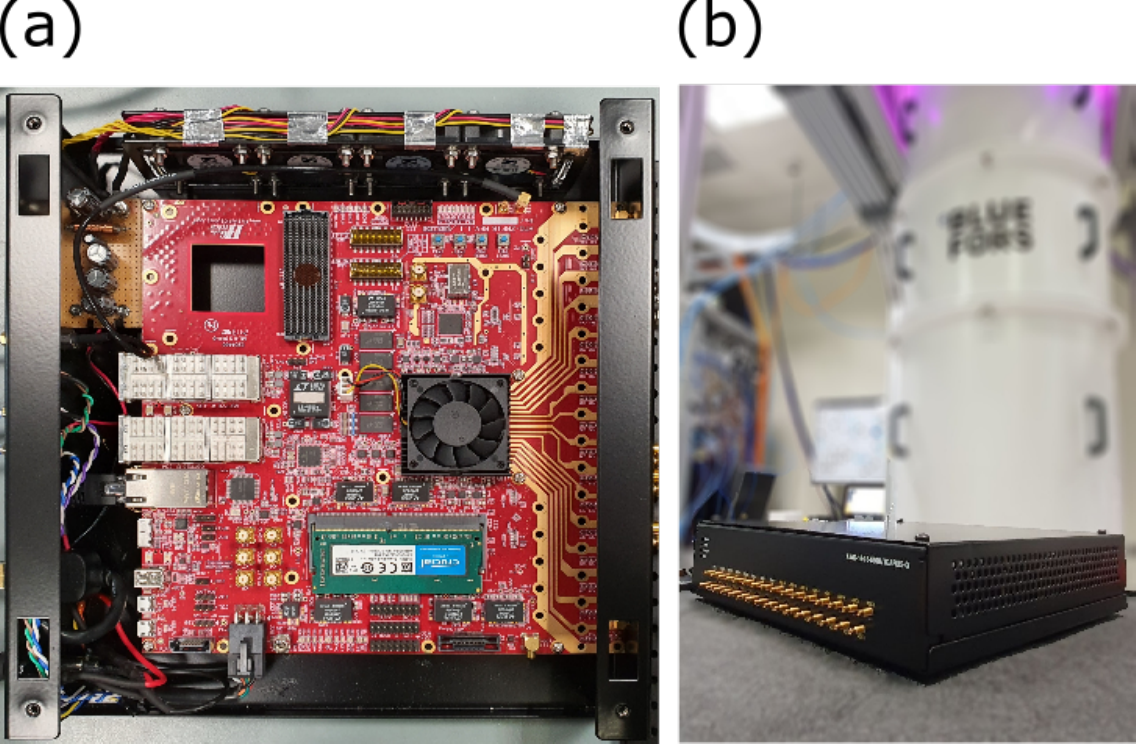}
	\caption{ICARUS-Q. (a) The RFSoC board (HTG-ZRF16), hosted within a casing, contains a unit of XCZU29DR RFSoC by Xilinx. (b) The front panel view of the enclosed RFSoC board.}
	\label{pic:board}
\end{figure}

The RFSoC board used in this work is available commercially (HTG-ZRF16, HiTech Global) and is populated with one unit of XCZU29DR RFSoC device (see Fig.~\ref{pic:board} for photo) that comes with sixteen 14-bit DAC (6.554~GS/s) and sixteen 12-bit ADC (2.058~GS/s) differential pins. 
On the RFSoC board, these are converted to single-ended signals using baluns (which support frequencies from 10~MHz to 8~GHz) and SSMC ports (support frequencies up to 12.4~GHz) for all of the DAC and ADC channels.  
Despite a rated sampling rate of 6.554~GS/s for the DACs and 2.096~GHz for the ADCs, our setup reaches a limit of 6.144~GS/s for the DACs and 1.96608~GS/s for the ADCs. This limit occurs because RFSoC FPGA is linked to the master clock and the sampling rates are multiples of the master clock (122.88~MHz).

In our setup, more DAC channels are needed than ADC channels. To allocate more of the limited FPGA block RAM memories~\cite{memoryprob} to the DACs, we reduced the number of active ADC channels to eight. 

\subsection{FPGA Logic}
\label{sec:FPGA}

\begin{figure}[!h]
	\centering
	\includegraphics[width=\linewidth]{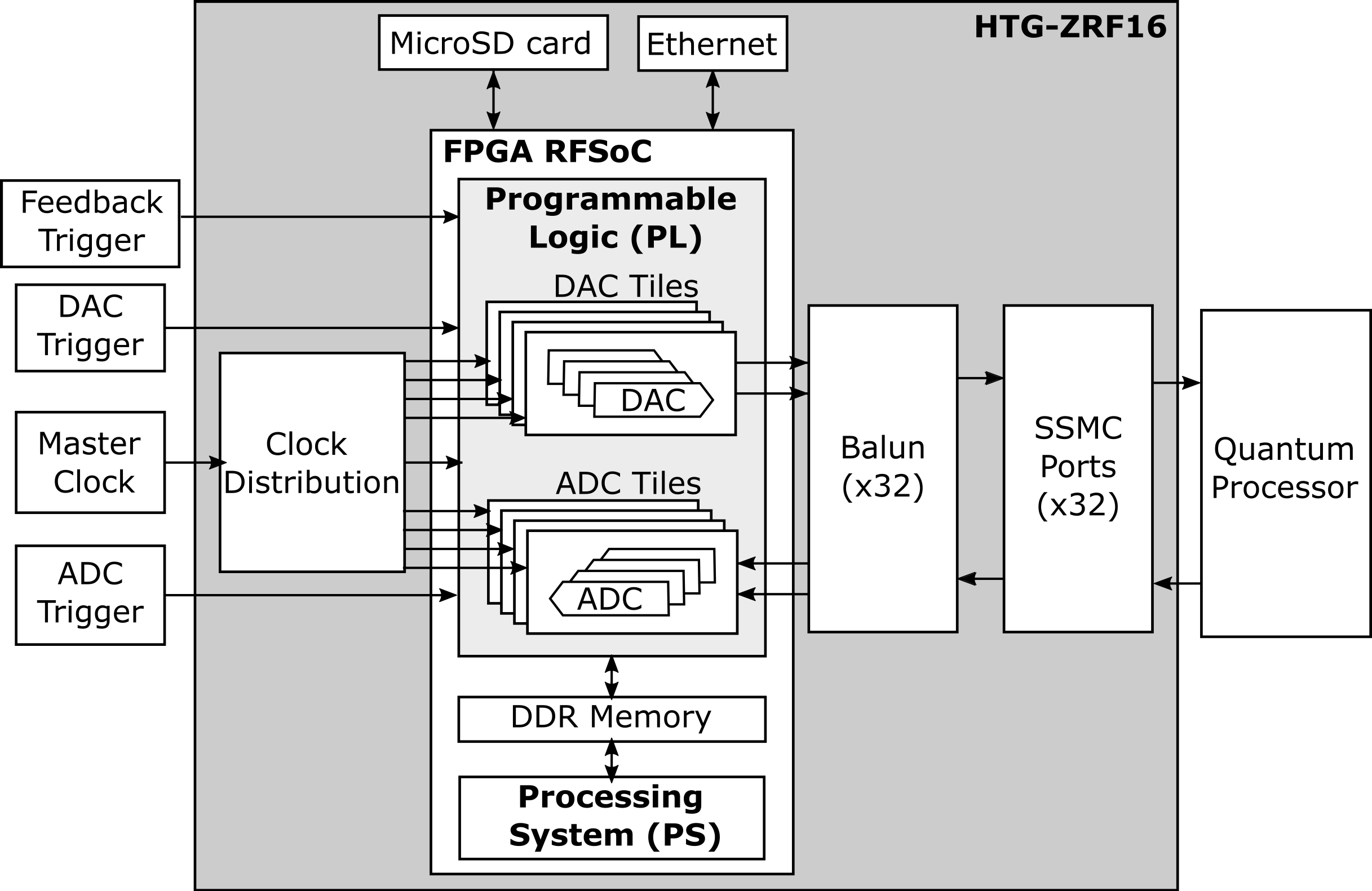}
	\caption{Some of the features on the HTG-ZRF16 board that are used in this setup. The board takes in multiple triggers and a clock signal as reference for the FPGA logic. The clock distribution subsystem distributes it to the FPGA as well as the DACs and ADCs tiles. The DACs and ADCs tiles generates differential microwave signals, which are converted to single-ended signals and transmitted out via SSMC ports. The board also supports communication via microSD card and Ethernet.}
	\label{fig:wholelogic}
\end{figure}

This section describes the FPGA logic that enables data movement between the PC and the quantum processor (see Fig.~\ref{fig:wholelogic}). 
The conversions between data and signals are performed by the ADCs and DACs, which are activated by the external triggers. For the current FPGA design with sixteen active DACs and eight active ADCs, we utilised around 50\% of the configurable logic blocks, 47\% of the total FPGA  RAM (BRAM usage is around 75\% without using Ultra RAMs) and almost none of the Digital Signal Processing (DSP) slices (0.12\%). This should leave space for future improvement specially on real time calculation on the acquired signals.
The HTG-ZRF16 also contains a two-stage clock distribution logic for phase synchronisation of the ADCs and DACs.
The FPGA also encompasses the Ethernet, microSD card and DDR memories on the HTG-ZRF16 board. In the following parts, we described the FPGA logics for the DAC and ADC implementations.

\subsubsection{FPGA Logic for DAC waveforms generation}

\begin{figure*}[!ht]
	\centering
	\includegraphics[width=\linewidth]{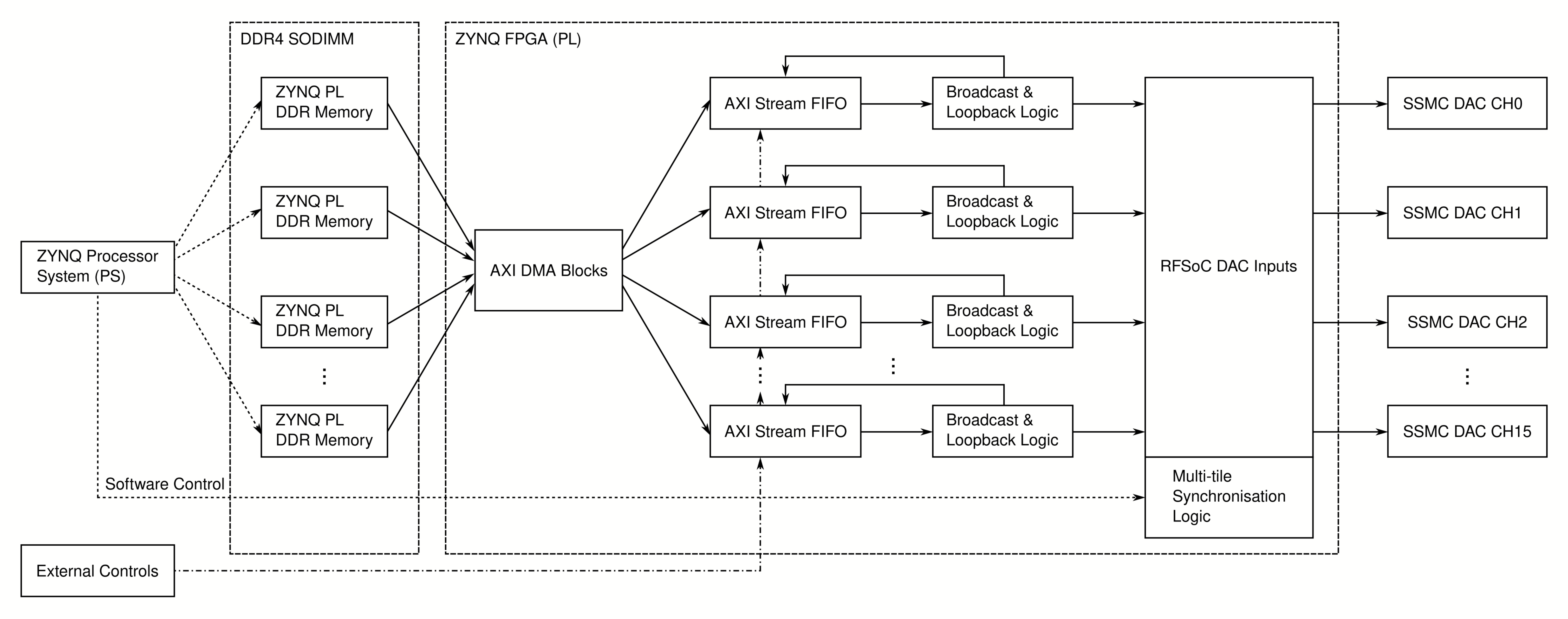}
	\caption{The FPGA logic for the DAC waveforms generation. The figure shows the data flow from the processor system (PS), going through the DDR memory and into the AXI Stream FIFO before it is sent out through the sixteen DAC channels. A multi-tile synchronisation logic ensures that all channels are synchronised.}
	\label{pic:daclogic}
\end{figure*}

The arbitrary waveforms are generated using RFSoC DAC (see Fig.~\ref{pic:daclogic}).
The sixteen DAC channels are powered by four DAC `tiles' inside the RFSoC chip. In order to ensure synchronous output of all DACs within a single board, multi-tiles synchronisation logic inside the RFSoC is utilised to calibrate the `tile-to-tile' time skew.

The DAC data flow starts from loading the waveform of interests into the programmable logic (PL) DDR memory. PL DDR memory is a physical SODIMM memory module connected to the PL-FPGA. The waveforms data are then moved into AXI Stream FIFO (First-In, First-Out) of each DAC channels. This is essential to allow the DAC playback to start simultaneously for all channels. However, due to the memory capacity limit of the internal block RAMs, individual AXI Stream FIFO can store up to 65,536 samples of DAC waveform of each channel.

The DAC waveform playback supports loopback function. When enabled, it allows the waveforms to be reloaded into AXI Stream FIFO without the need of reloading the waveform data from a host computer, thus reducing the overhead time of re-arming the DAC for next DAC playback.

Once the waveforms data are loaded into AXI Stream FIFO of each channel, the system will wait for an external trigger event from the external control logic before starting the DAC waveform playback. The trigger signal from external control logic applies to all DAC channels so the output can be streamed out via the SSMC connectors simultaneously. The external control logic also supports the waveform data swap. When enabled, the waveforms of upper eight channels can be swapped to the lower eight channels to support more advanced pulse sequences (see Sec.~\ref{sec:feedback} for more information).

\subsubsection{FPGA Logic for ADC waveforms acquisition}

\begin{figure*}[!ht]
	\centering
	\includegraphics[width=\linewidth]{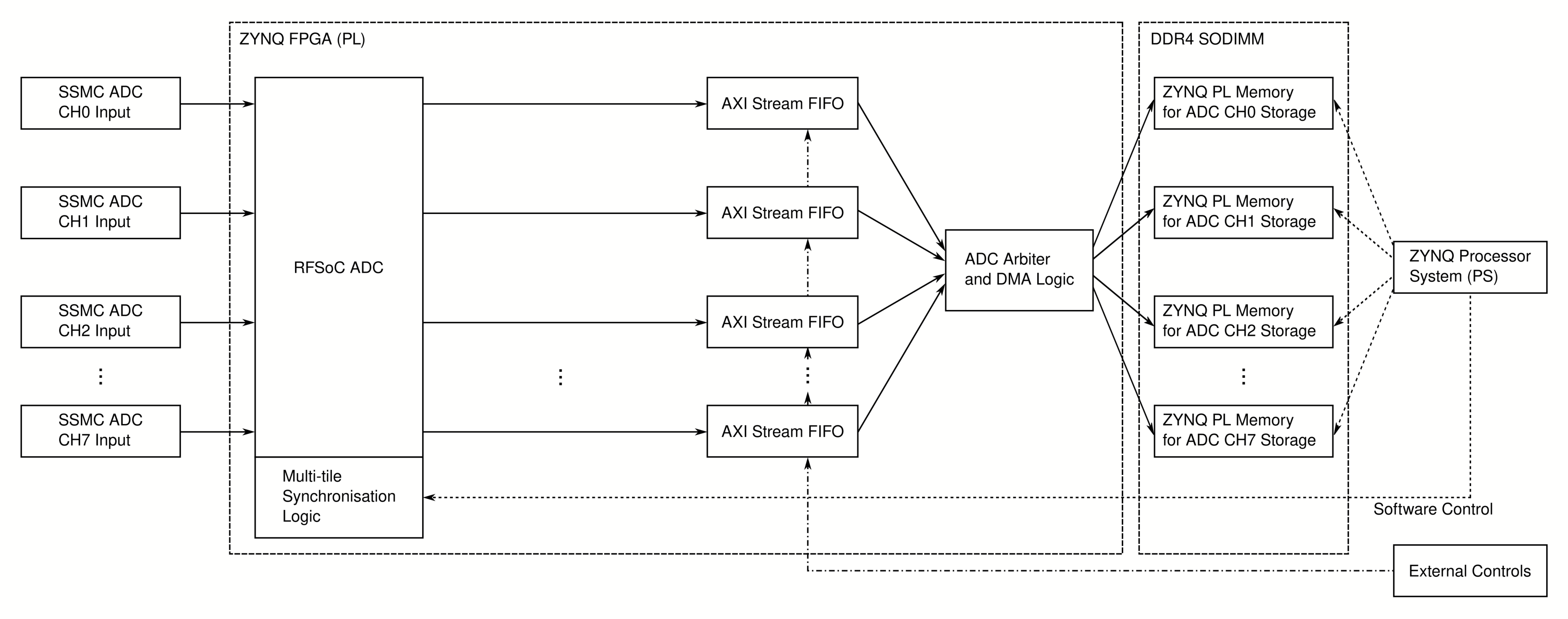}
	\caption{The FPGA logic for the ADC waveforms acquisition. The data from the eight ADC channels are streamed to the AXI Stream FIFO gets moved into the DDR memory via the ADC Arbiter and DMA Logic, only when a trigger is received. The ADC arbiter and DMA logic then transfers the data to the programmable logic (PL) memory for storage when an instruction is received from the processor system (PS). A multi-tile synchronisation logic ensures that all channels are synchronised.}
	\label{pic:adclogic}
\end{figure*}

In our firmware, the waveform acquisition system is powered by eight ADC channels of the RFSoC (see Fig.~\ref{pic:adclogic}). The analog input is fed into RFSoC ADC via SSMC connectors. ADC digitise the incoming waveform continuously, but is not streamed to the AXI Stream FIFO without the external ADC trigger.

When the external ADC trigger event occurs, 65,536 samples of digitised waveform data for each ADC channel will be stored in the FPGA AXI Stream FIFO. The data stored in AXI Stream FIFO is moved to the external DDR4 SODIMM, awaiting subsequent process/instruction from ZYNQ Processor System (PS). The data can be stored in both HEX or ASCII file format depending on the applied settings for subsequent analysis. The ADC trigger will be re-armed after the acquired data has completely the transfer to the ZYNQ Processor System.

\subsection{Microwave Generation \& Detection}
\label{sec:microwave}
The RFSoC has gained some degree of attention and its performance has been tested by several groups~\cite{kalfus2020high, liu2021characterizing}. 
The on-board DACs output arbitrary waveform, generated from 65,536 samples at variable sampling rates, up to 6.144~GS/s. At its maximum sampling rate, this translates to about 10~$\mu$s of waveform points. The ADC also stores equal number samples but operates at 1.96608~GS/s, resulting in a waveform of about 33~$\mu$s. 
After triggering the DAC channels, there is a minimum delay of about 30~$\mu$s before triggering the next pulse. 
To further evaluate the performance of the DACs and the ADCs, we performed several tests and describe their results in the following.

\subsubsection{Arbitrary Waveform Generation}
\label{sec:awg}
\begin{figure}[!h]
	\centering
	\includegraphics[width=\linewidth]{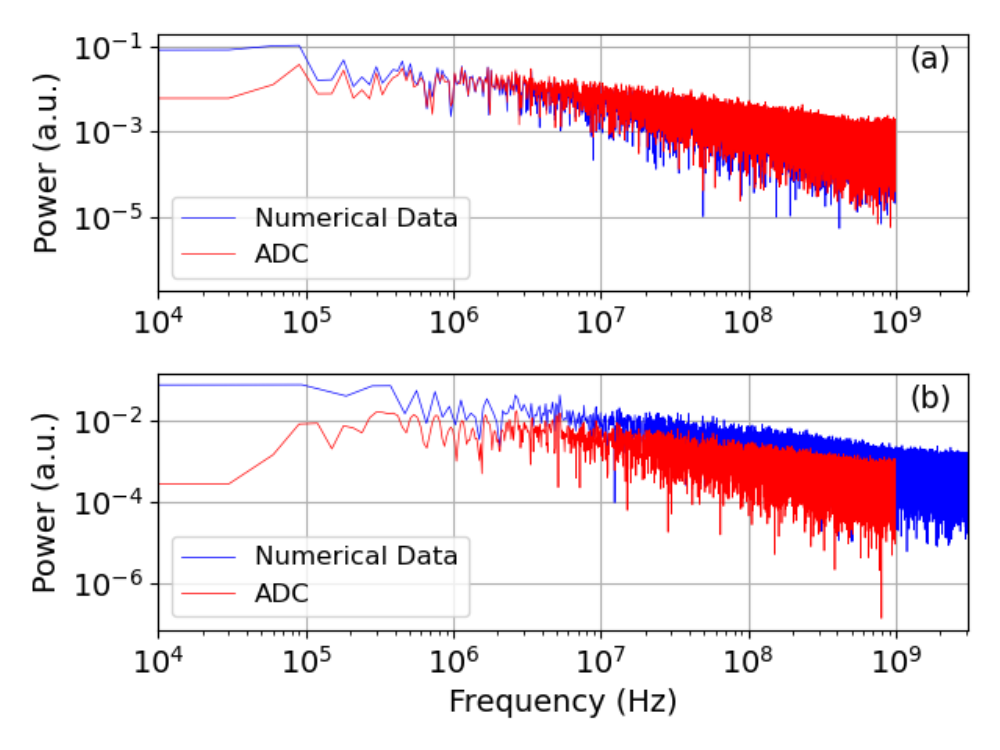}
	\caption{Pink noise generated by the DAC at (a) 1.96608~GS/s and (b) 6.144~GS/s. 
	The pink noise waveform was calculated using the Voss algorithm~\cite{voss} and the DACs were fed back directly into the ADCs without any filters for sampling. 
	For both runs, the generated waveform was sampled with the ADC at 1.96608~GS/s. All plots stop at their respective Nyquist frequencies, which is one half of the sampling frequency.}
	\label{fig:pinknoise}
\end{figure}

In a typical quantum computer experiment, rectangular or Gaussian-shaped pulses are common but pulses with arbitrary phases and amplitudes~\cite{motzoi2009simple, economou2015analytical} are also often used.
Some quantum information processing applications also requires the use of non-gate-based signals such as  optimal control theory~\cite{khaneja2001time, tabuchi2017design, werninghaus2021leakage}, adiabatic quantum computation, continuous variable quantum computing~\cite{song2017continuous}, etc.
In our setup, the modulated pulses are numerically designed and generated up to 3.072 GHz (one half of the DAC's maximum sampling rate).
To demonstrate true arbitrary waveform generation capabilities, we tested the DAC with pink noise and compared the generated signal against the calculated waveform datapoints (see Fig.~\ref{fig:pinknoise}). 
The pink noise waveform was calculated using the Voss algorithm~\cite{voss} and the signal was generated by the DAC at two different sampling rates: 1.96608~GS/s and 6.144~GS/s. Both waveforms were sampled using the ADC at 1.96608~GS/s. 

In Fig.~\ref{fig:pinknoise}(a), the DAC samples and the ADC data are plotted in frequency domain, respectively. The frequency profiles bear a qualitative resemblance with each other across the frequency components, except below 1~MHz. We attribute this to the balun's supported frequency range (10~MHz--8~GHz) where frequencies below 10~MHz are attenuated, akin to a high pass filter.


\subsubsection{Nyquist Zone Implementation}
The Shannon-Nyquist sampling theorem states that a signal can be adequately generated or sampled at frequencies below half of the sampling rate. This frequency threshold is known as Nyquist frequency. 
However, generating or sampling signals in discrete-time creates aliases that are mirrored repeatedly across multiples of the Nyquist frequency~\cite{soliman1990continuous, kalfus2020high}.  Each ``segment'' of the frequency domain is commonly referred to as Nyquist zones.
With careful planning, one can utilise frequencies above the first zone without upgrading existing electronics~\cite{erdmann201716, kalfus2020high, stefanazzi2021qick}.

For controlling the qubits, we digitally generate shaped sine wave pulses directly in the GHz frequency range and use them in their respective Nyquist zones, which also naturally preserves the phase coherence of the qubit. As such, this method does not require a local oscillator nor its modulation. 

However, there are implications for using this approach. The voltage in time domain, $v(t)$ described by:
\begin{equation}
    v(t) = \left[x(t)\sum_{k=-\infty}^\infty \delta(t-kT)  \right] \ast  r(t),
\end{equation}
\noindent is affected by the ``reconstruction waveform'', $r(t)$~\cite{kalfus2020high}. Here, $x(t)$ is the analytical function we sample, $\delta(t-kT)$ is the Dirac delta function, and $T$ is the sampling period. The Fourier transform of $v(t)$ is:
\begin{equation}
    V(\omega) = \left[ X(\omega) \ast \sum_{n=-\infty}^\infty\delta(\omega T - 2 \pi n) \right] \times R(\omega).
    \label{eq:nyq_freq}
\end{equation}
\noindent where, $X(\omega)$ is the Fourier transform of $x(t)$, and $R(\omega)$ is a sinc function that is determined by the DACs operational mode~\cite{rfsocDatasheet}. The RFSoC used here supports two modes: the non-return-to-zero (NRZ) mode and the mixed mode. The respective reconstruction waveforms in the Fourier space are two different sinc functions:
\begin{equation}
    R_{\mathrm{NRZ}}(\omega) = T e^{-i\omega T/2}  \mathrm{sinc} \left( \frac{\omega T}{2}\right)
    \label{eq:nrz}
\end{equation}
and
\begin{equation}
    R_{\mathrm{mix}}(\omega) = \frac{\omega T^2}{4} e^{-i(\omega T - \pi)/2}  \mathrm{sinc}^2 \left( \frac{\omega T}{4}\right)
    \label{eq:mix}
\end{equation}
respectively.

Therefore, when using the ADC with its higher Nyquist zones, we expect the signal-to-noise ratio (SNR) to degrade to an extent. We investigated this by using the ADC at 1.96608~GS/s to sample various signals generated at higher frequencies (zones) which were aliased to 800~MHz within the first zone of the ADC.  
The DAC decoder mode was set to normal (NRZ) mode for this test.

\begin{figure}[!h]
    \centering
	\includegraphics[width=\linewidth]{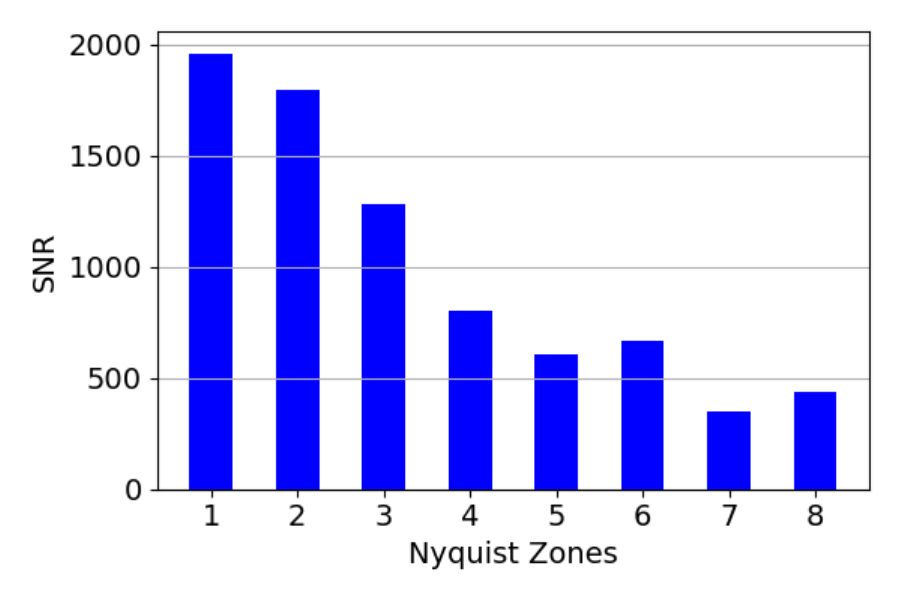}
	\caption{\red{SNR measurements of DAC output at different ADC Nyquist zones. The signal in the first zone is at 800 MHz and the ones in the higher zones are its aliases. The DAC output is passed through filters to only select the alias frequency that corresponds to each ADC Nyquist zone. For example: The DAC outputs a 4732.16 MHz signal (using its second zone) and is filtered with a 2.7-6.0 GHz bandpass filter; it is sampled in the ADC's as 800 MHz (using its fifth zone). The SNRs were calculated in frequency domain using the signal peak amplitude and the average noise within a bandwidth of 100 MHz.}}
	\label{pic:snr}
\end{figure}

The 800~MHz signal from the RFSoC DAC, in the first Nyquist zone of the ADC sampling rate, is measured to have $\text{SNR} \approx 2\times{}10^{3}$ (see Fig.~\ref{pic:snr}). The 7.06432~GHz signal, the 8$^{\text{th}}$ Nyquist zone alias of the 800~MHz, is in the similar range of the typical qubit transition frequency. At this frequency, the SNR is measured to be around $4\times{}10^{2}$, which is around 5 times lower than when using 800~MHz.

\subsubsection{Power-Frequency Dependence}
\begin{figure}[!h]
    \centering
	\includegraphics[width=\linewidth]{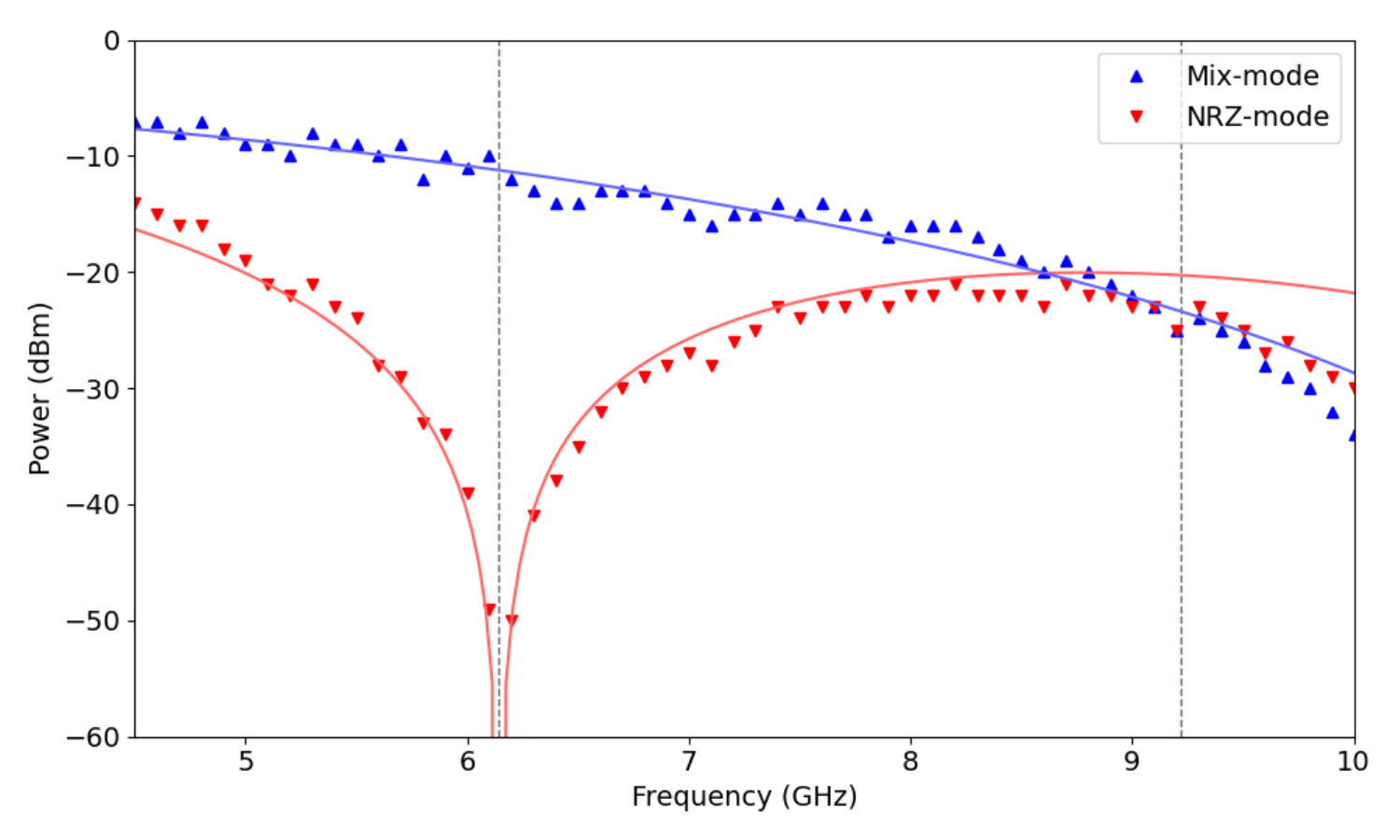}
	\caption{Output power of the RFSoC DAC from 4.5 to 10~GHz, corresponding to parts of the first three Nyquist zones. The zones are delineated by the dashed lines.}
	\label{pic:powerlinearity}
\end{figure}

Our approach to use the higher Nyquist zone involves a power dependency as defined in Eq.~\ref{eq:nyq_freq}. Since superconducting qubits are controlled and readout in the frequency domain at unique frequencies, we decided to investigate its output power-frequency dependence, particularly in between the DAC Nyquist zones. We measured the DACs output power using a spectrum analyzer from 4.5~GHz to 10~GHz, which is around the typical range for superconducting qubits and their readout resonators~\cite{arute2019quantum, krinner2019engineering}. The results are presented in Fig.~\ref{pic:powerlinearity} where Eqs.~\ref{eq:nrz} and \ref{eq:mix} were fitted to the measured DAC output power for the normal (NRZ) and the mix modes respectively.

For the normal (NRZ) mode, the output power dipped at 6.144~GHz as expected. Between 7--10~GHz, the power averaged at $-24.1 \pm 2.4$~dBm, which improved slightly between 7--9~GHz (in terms of standard deviation) to $-23.1 \pm 1.8$~dBm. Although the power variation is higher compared to microwave synthesizers or high-end benchtop AWGs, these error margin are not expected to pose significant problems since the qubits would be characterised/calibrated periodically at fixed frequencies. For the mix-mode, the expected power dip takes place at 12.288~GHz (double the sampling rate). The average power for this mode was at $-20.4 \pm 5.7$~dBm between 7--10~GHz, and $-16.9 \pm 2.3$~dBm between 7--9~GHz.

The output power was observed to deviate downwards from the fitted plots at higher frequencies. We attribute this observation to the on-board balun's supported frequency range (10~MHz--8~GHz), beyond which some attenuation is to be expected, similarly to the power spectra in Sec.~\ref{sec:awg}---except acting as a low pass filter here.

\subsection{Multi-channel and Multi-board Operation}
\label{sec:sync}
In order to scale up the number of DAC and ADC channels used to control and measure the qubits, the channels need to be able to output the waveforms at a synchronised timing and phase. 
There are two kinds of synchronisation we ought to achieve: (1) intra-board inter-channel synchronisation, and (2) inter-board synchronisation.

\begin{figure}[!h]
	\centering
	\includegraphics[width=\linewidth]{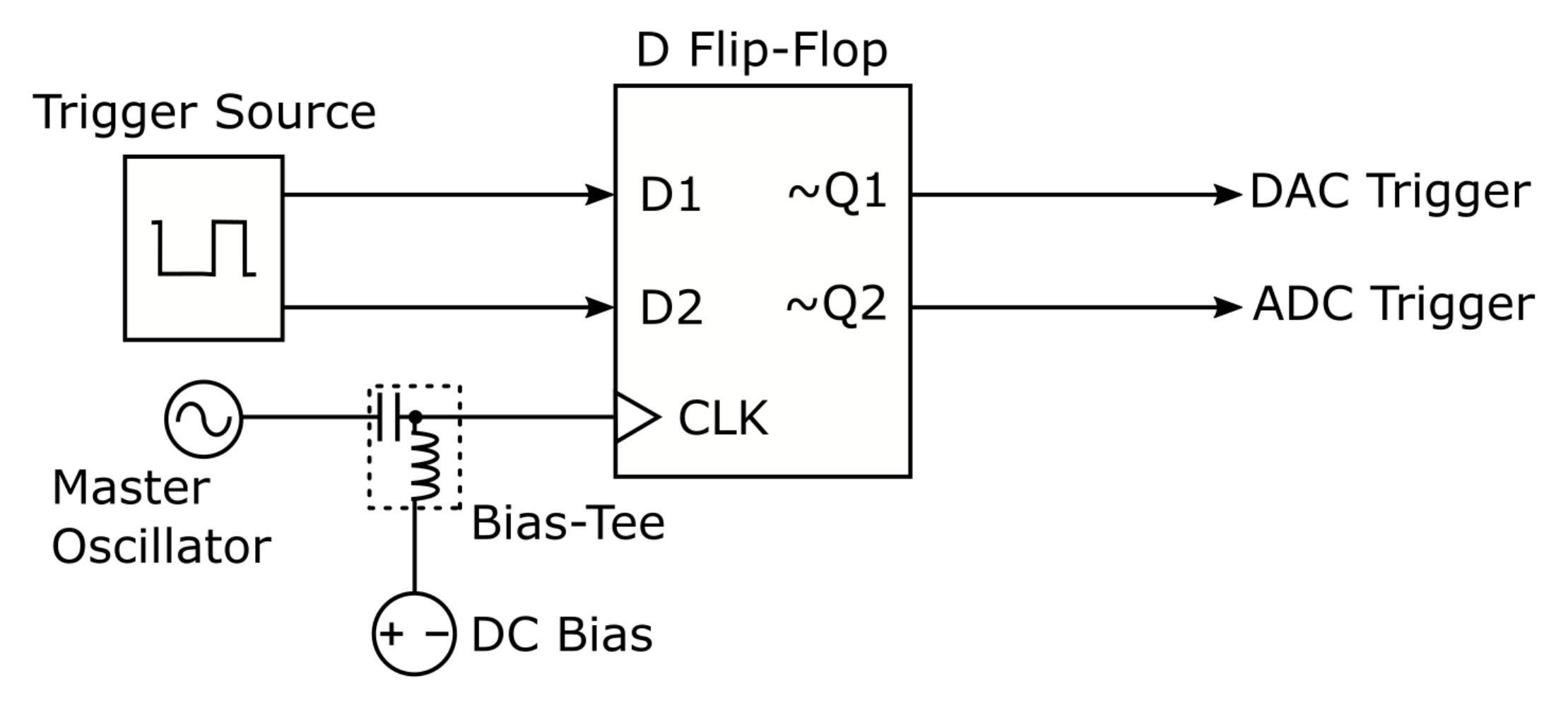}
	\caption{Schematic of the d-type flip-flop circuit. 
	The two independent falling-edge trigger signals for the DAC outputting and the ADC sampling (Q1 and Q2) are provided by the trigger source (D1 and D2) and are synchronised to a master oscillator using a dual-channel d-type flip-flop. 
	An appropriate DC bias is supplied to the clock input of the flip-flop, so that the AC clock signal from the oscillator moves through the on/off levels.}
	\label{fig:flipflop}
\end{figure}

Within a RFSoC board, the inter-channel synchronisation is achieved through the multi-tile synchronisation (MTS) logic in the firmware, which utilises the on-board phase-locked loop (PLL) to lock the channel outputs to the external reference clock. The inter-board synchronisation, on the other hand, is achieved by carefully distributing the single master oscillator to all of the boards such that they have the same reference clock signal for synchronisation. To generate and sample the waveforms via the DACs and the ADCs,  in a synchronised fashion, hardware triggers are implemented. 
To prevent metastability of sampling the external trigger by the RFSoC, the trigger signals are first synchronised to the master clock signal using an external d-type flip-flop (see Fig.~\ref{fig:flipflop}). The synchronised triggers are then distributed to multiple RFSoC boards.  This ensures that every RFSoC board samples the trigger at the same clock cycle~\cite{intelMeta}.

With the master clock, MTS and synchronised triggers, phase coherence between pulses was preserved for all channel across with multiple boards without the use of any local oscillator. 

\begin{figure}[!h]
	\centering
	\includegraphics[width=\linewidth]{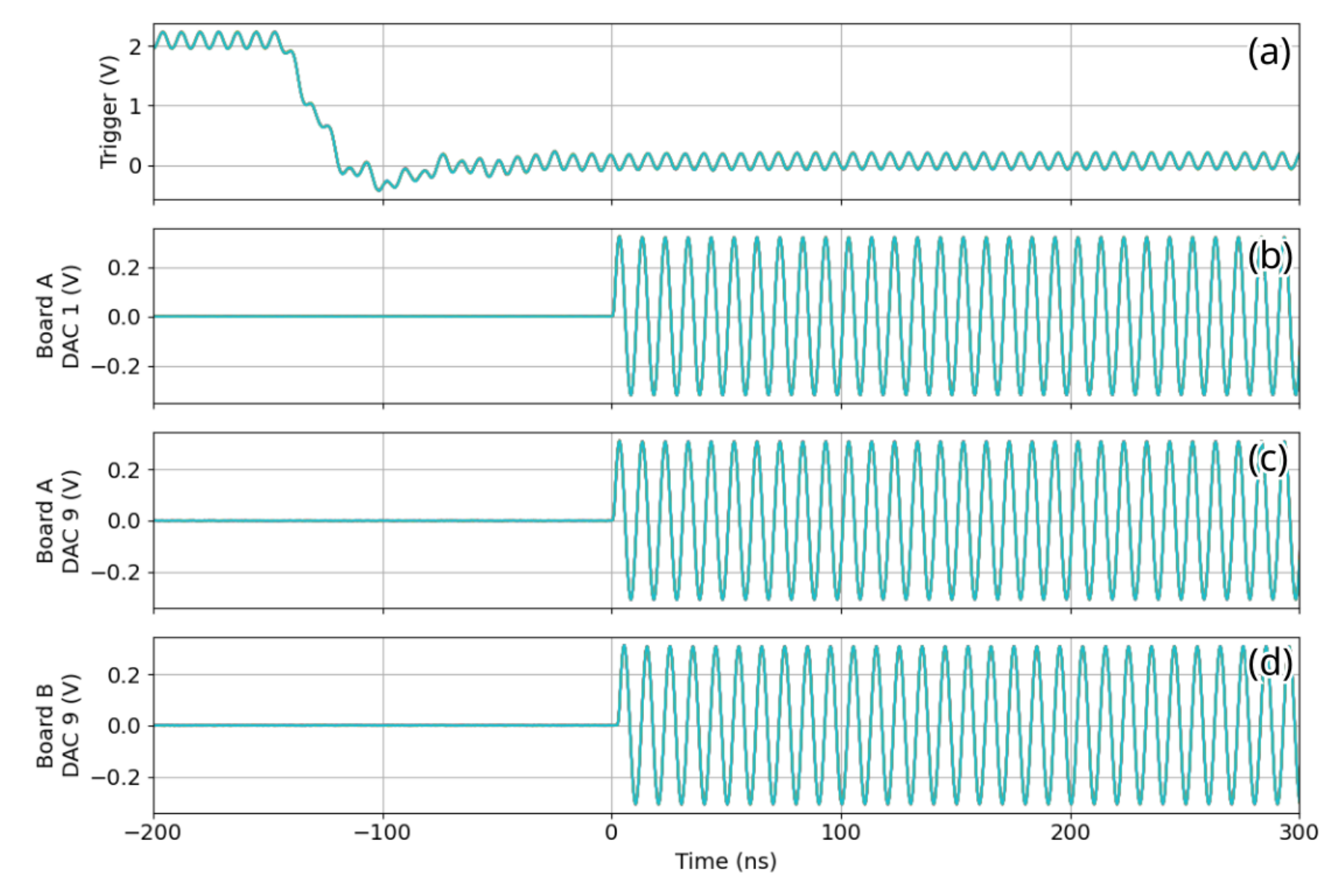}
	\caption{Synchronisation of the DAC channels from multiple boards. (a) Using a synchronised to the trigger signal, 10 runs of the DAC were captured. (b-d) The DAC outputs from two RFSoC boards (Boards A and B) were connected to an oscilloscope and synchronisation between channels of the same board and between multiple boards were achieved.}
	\label{fig:multiboard}
\end{figure}

Using the MTS logic and the trigger signal through the d-type flip-flop, the DAC outputs from two RFSoC boards were synchronised (see Fig.~\ref{fig:multiboard}). The oscillations in the trigger signal are caused by the leakage of the clock signal through the flip-flop, but did not affect the trigger reception by the RFSoC boards. The slight delay on Board B is caused by the difference in the lengths of the trigger distribution paths - which can be easily corrected by exactly matching the cable lengths (or by introducing delays in the software). 

\subsection{Feedback Control}
\label{sec:feedback}

\begin{figure}[!h]
	\centering
	\includegraphics[width=\linewidth]{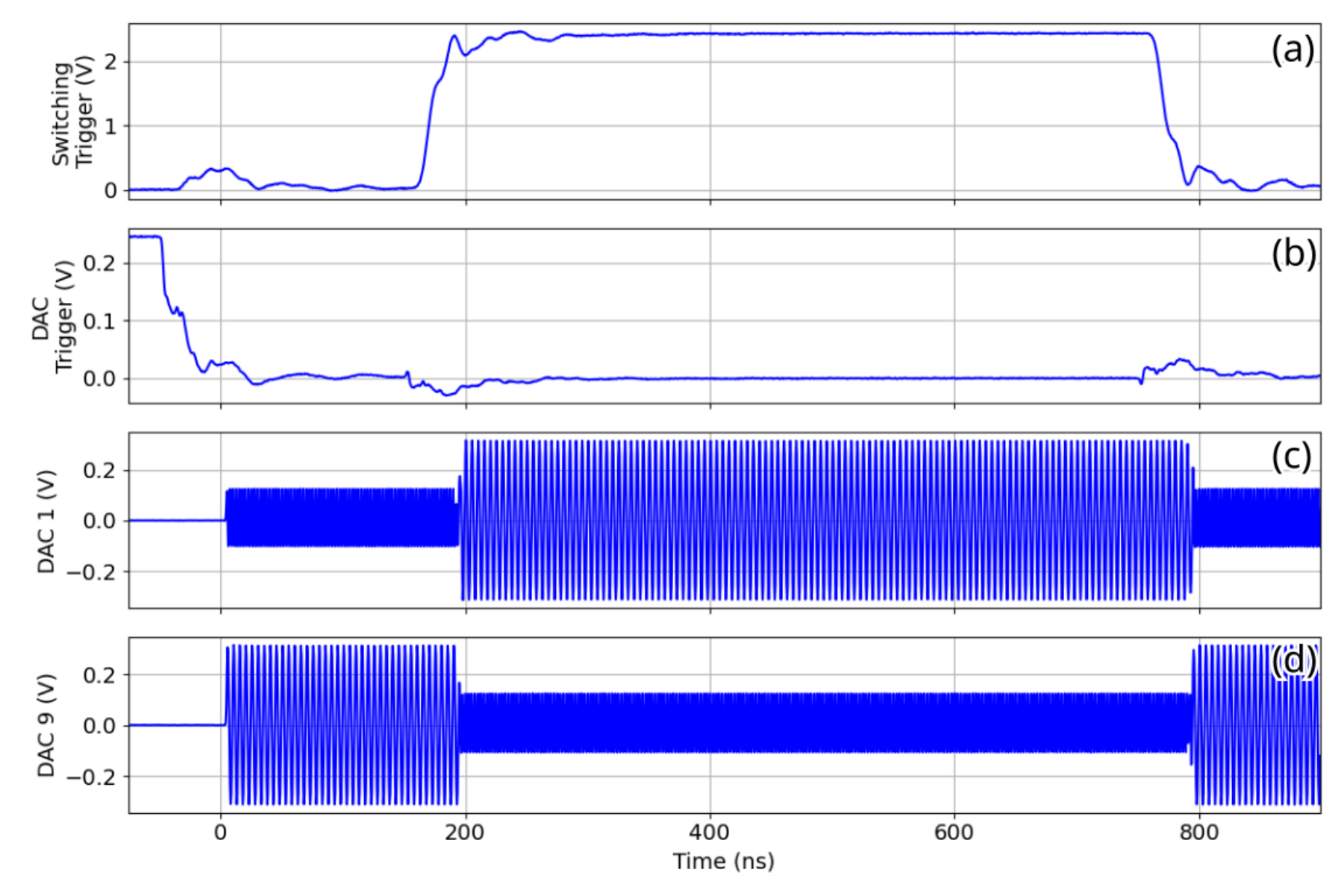}
	\caption{DAC playback switching with hardware trigger signals. (a) The switching trigger sent to the RFSoC. (b) The DAC trigger sent to the RFSoC. (c-d) Upon receiving the switching trigger, the upper DAC memory banks (DACs 0 to 7) switches with  the lower banks (DACs 8 to 15). The delay from the trigger to completion of switching of the channels took approximately $20 \pm5$ ns.}
	\label{fig:feedback}
\end{figure}

Having the ability to switch the waveform in real-time (nanosecond-scale) allows for the possibility to correct the qubit state in the midst of running quantum circuits. The RFSoC integrates waveform switching based on hardware trigger signal. Upon receiving the switching trigger, the outputs of the DAC channels in the upper memory banks (0 through 7) switches to those in the lower memory banks (8 through 15) within a few nanoseconds (see Fig.~\ref{fig:feedback}). 

\subsection{Measurements with Superconducting Qubit}
\label{sec:qubit}

The control system was used for experiments with a non-tunable superconducting qubit to demonstrate its capabilities. 
For the following qubit measurements, the low noise bias circuit described in Section~\ref{sec:dcbias} was not used.
We also upgraded the trigger source (PulseBlaster PB12-100-4K-PCIe) to receive the master clock directly without the need to use the flip-flop circuit.
The qubit is a transmon with a single Josephson junction on silicon substrate, placed in a 3D aluminium cavity. Two DAC channels, for driving the qubit and the cavity respectively, were combined through a microwave combiner and fed through the input port of the cavity. The readout port of the cavity was connected to an ADC channel. 

\subsubsection{Cavity Power Sweep}

\begin{figure}[!ht]
	\centering
	\includegraphics[width=\linewidth]{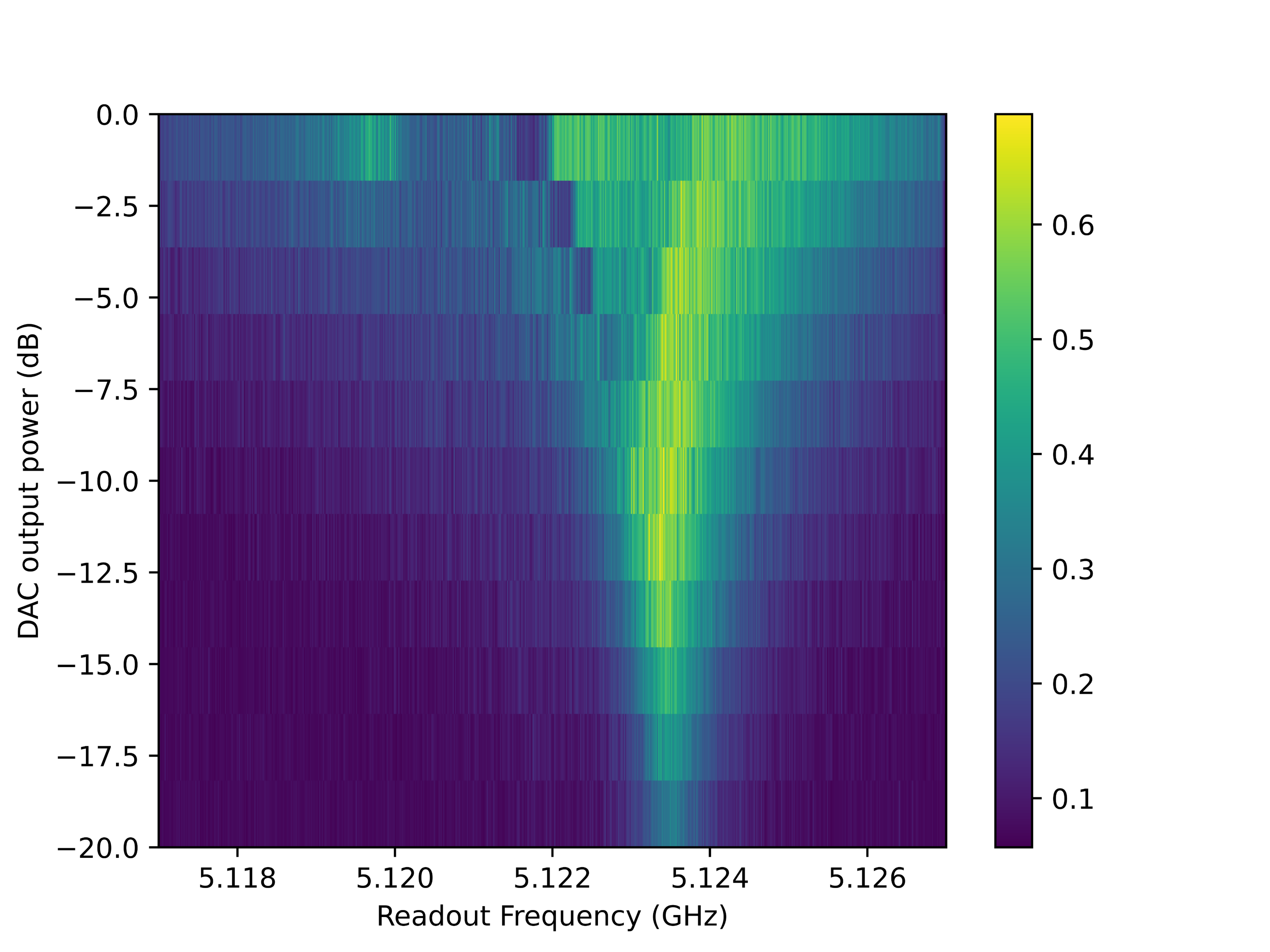}
	\caption{Cavity transmission response. We plot the amplitude of the Fourier transform of the readout signal from the ADC as a function of DAC power and cavity driving frequency. The cavity was driven for 10~$\mu$s near its resonance frequency over 20~dB of power with a rest period of 200~ms. At each point, the amplitude was averaged over 100 shots. The dressed cavity frequency is at 5.12375~GHz and from the data we are able to resolve the frequency at -12~dB. The bare frequency of 5.12~GHz is slightly visible at the highest power of the instrument.}
	\label{fig:powersweep}
\end{figure}

We first demonstrate the ability of the system to perform qubit readout by driving a readout signal through the cavity and measuring the response as shown in Fig.~\ref{fig:powersweep}. A range of readout frequencies from 5.117~GHz to 5.127~GHz were swept to test the frequency precision of the device. The sampling rate of the DAC was set at 5.89824~GS/s and the ADC was set at 1.96608~GS/s. 
These sampling rates were chosen based on the cavity and qubit frequencies and the corresponding output powers at different Nyquist zones. For this experiment, the DAC is operating at the second Nyquist zone and the ADC at the fifth Nyquist zone. 

Furthermore, we test the voltage control of the instrument by varying the drive power to resolve the dispersive shift of the cavity without the use of RF attenuators. While the maximum driving power of the device was unable to fully resolve the bare frequency of the cavity, we are able to resolve the dispersive shift of 3.8~MHz and the dressed cavity frequency at -12~dB.

\subsubsection{Rabi Spectroscopy}

\begin{figure}[!ht]
	\centering
	\includegraphics[width=\linewidth]{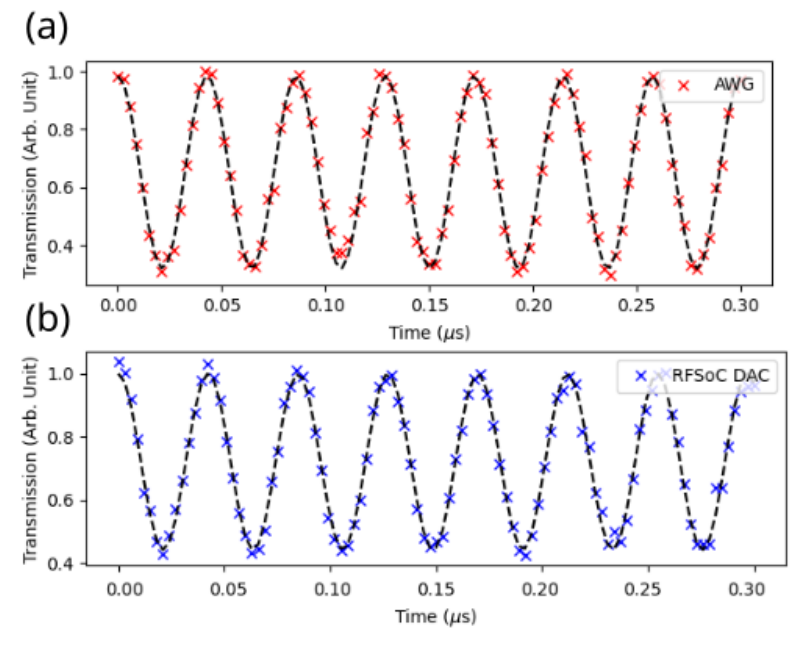}
	\caption{Rabi oscillation of the qubit using (a) AWG and (b) RFSoC DAC, respectively. The qubit was driven with pulse durations up to 300~ns in 3~ns intervals. Each point was averaged over 1000 measurements. Through a sine wave fit, we approximate the Rabi oscillation frequency $f_{\text{Rabi}}$ as 23.3~MHz for the AWG and 23.5~MHz for the RFSoC. We also obtain the $\pi$-pulse duration $t_{\pi}$ of 21.5~ns for the AWG and 21.2~ns for the RFSoC.}
	\label{fig:rabi}
\end{figure}

Next, we move on to demonstrating qubit control with the RFSoC. We do this with Rabi spectroscopy, varying the length of the qubit driving pulse and measuring the cavity response (see Fig.~\ref{fig:rabi}). The qubit was driven at its resonance frequency of 3.357~GHz on the second Nyquist zone on DAC channel 1, tile 1 and measured with a readout pulse of length 5~$\mu$s on DAC channel 13, tile 4. We also compare this to a separate experimental setup AWG (described in Ref.~\cite{gauss_ghost_factors}).

From the oscillation, we are able to drive and resolve the 3~ns step rotations of the qubit. Hence, we demonstrate the ability of the RFSoC to perform and measure arbitrary rotations of the qubit and synchronise DACs on separate tiles.

\subsubsection{Ramsey Spectroscopy}

\begin{figure}[!ht]
	\centering
	\includegraphics[width=\linewidth]{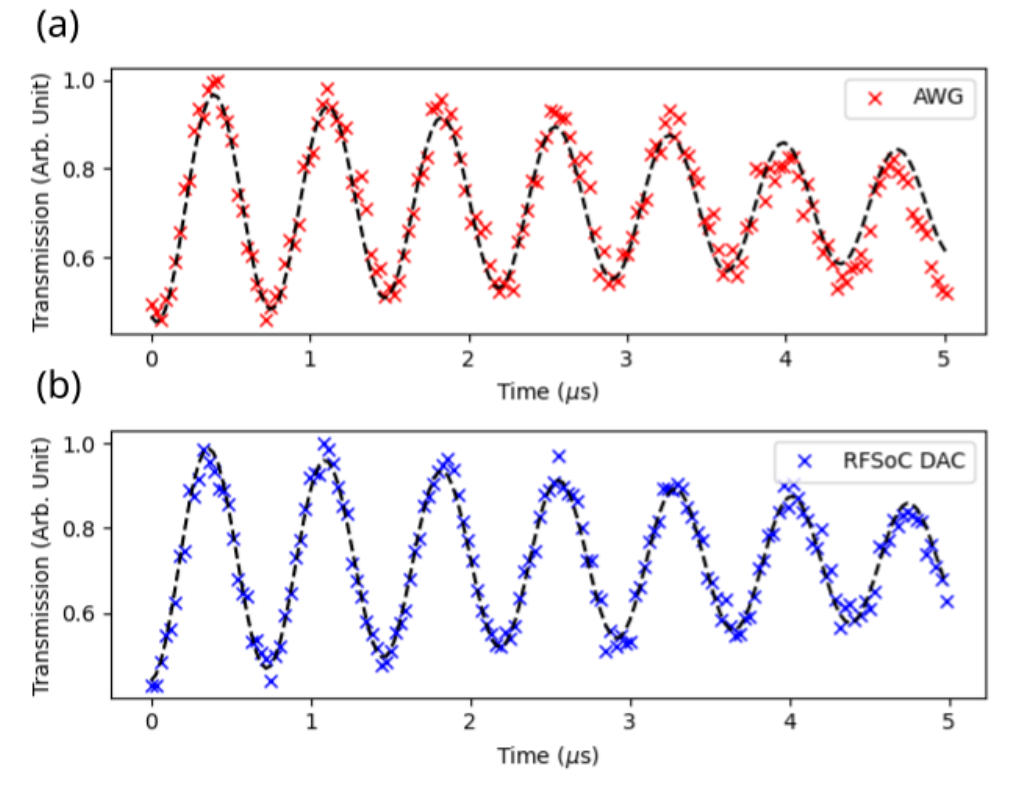}
	\caption{Ramsey fringes. We vary $\tau$ up to 5~$\mu$s and a apply a frequency detuning of 1~MHz using (a) the AWG and (b) the RFSoC. Each point was averaged over 1000 measurements. The data is fitted to a sine wave with an exponential decay, yielding an oscillation of frequency 1.367~MHz and dephasing time of 6.66~$\mu$s for the RFSoC and similar values of 1.38~MHz and 6.52~$\mu$s for the AWG, respectively.}
	\label{fig:ramsey}
\end{figure}

We further test the phase stability of the system over half of the available DAC playback time through a Ramsey experiment (see Fig.~\ref{fig:ramsey}). We apply two $\pi/2$-pulses separated by flight time $\tau$ before readout. 
From the Ramsey fringes, we do not observe significant deviations compared to AWG in 5~$\mu$s of free evolution time. Hence, the system appears to be stable during full playback. 

\begin{figure}[!ht]
	\centering
	\includegraphics[width=\linewidth]{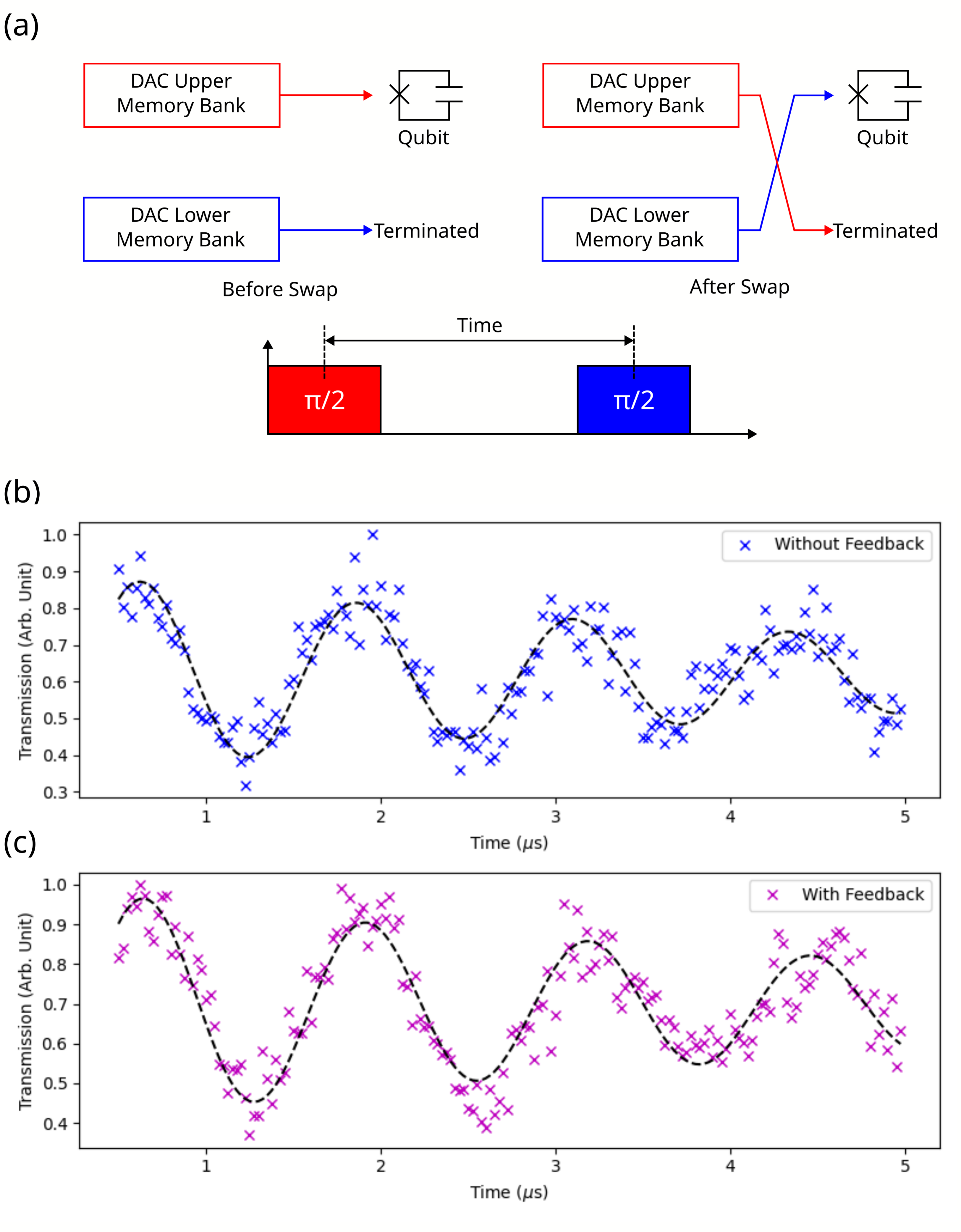}
    \caption{\red{Ramsey fringes performed with the feedback control (i.e.~swapping of the DAC memory banks). (a) To test the effects of the feedback, we carried out a Ramsey experiment where the first $\pi/2$-pulse originated from the original memory bank and the second $\pi/2$-pulse was generated from the swapped the memory bank.  The Ramsey fringes (b) without and (c) with feedback control. Each point was averaged over 1000 measurements.} The fits yield dephasing times of 4.84~$\mu$s and 5.08~$\mu$s, respectively. This experiment was performed at a different cooldown as compared to Fig.~\ref{fig:ramsey}.}
	\label{fig:ramsey_fb}
\end{figure}

Furthermore, we perform Ramsey experiment with the feedback control (see Fig.~\ref{fig:ramsey_fb}) to swap the waveform memory banks as described in Section~\ref{sec:feedback}. The switching of the memory banks is carried out in between the two $\pi/2$-pulses of the Ramsey sequence. Similarly, we do not observe significant deviations when using the feedback control for the Ramsey experiment.

\subsubsection{Quantum State Tomography}

\begin{figure}[!ht]
	\centering
	\includegraphics[trim={1.5cm, 2.5cm, 1.5cm, 1cm} , clip, width=0.7\linewidth]{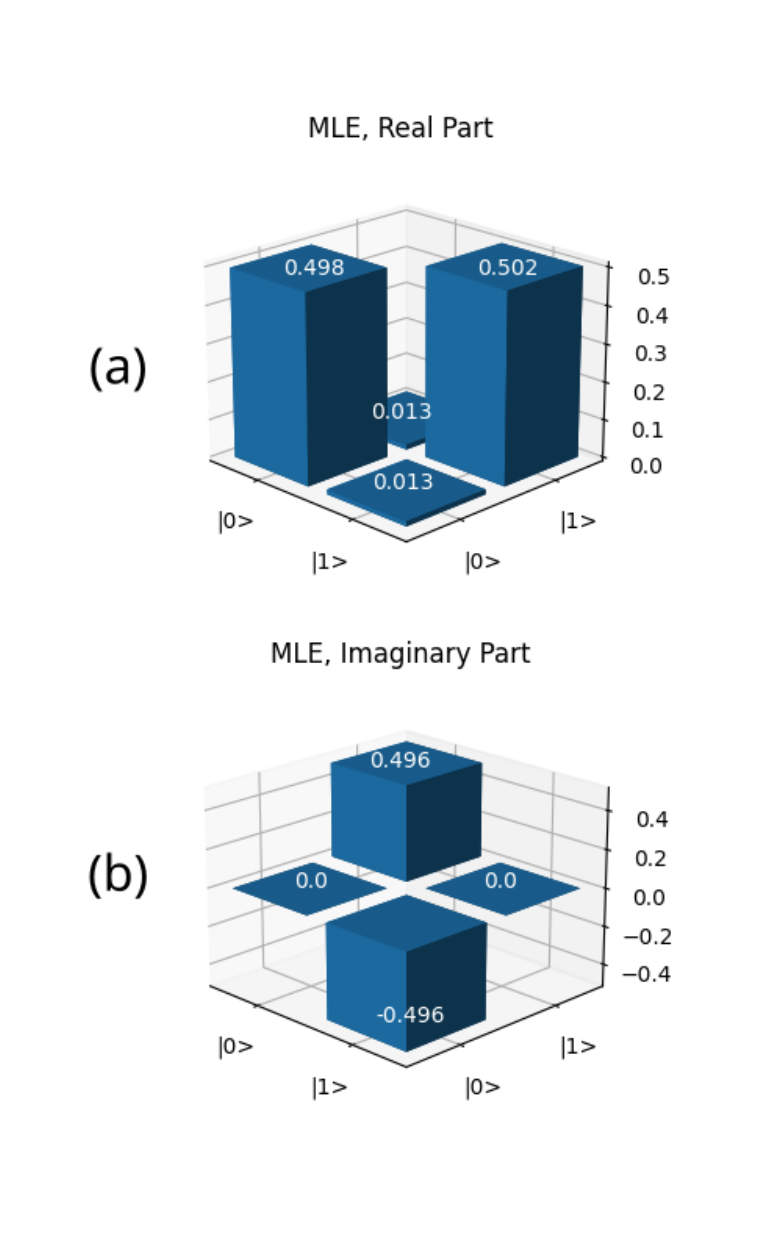}
	\caption{Quantum State Tomography for $\frac{1}{\sqrt{2}}(\ket{0} - i\ket{1})$. The figure shows the (a) real and (b) imaginary parts of the Maximum Likelihood Estimate (MLE) fit. A $\pi/2$-pulse is applied before prerotation. The state fidelity from MLE is $99.8\%$.}
	\label{fig:tomography01}
\end{figure}

Lastly, using $t_{\pi}$ and $t_{\pi/2}=t_{\pi}/2$ obtained from the results of Rabi spectroscopy, we perform Quantum State Tomography (QST) with Maximum Likelihood Estimate (MLE) using $I, X, X/2, Y/2$ as pre-rotation gates for the $\frac{1}{\sqrt{2}}(\ket{0} - i\ket{1})$ state to evaluate the state fidelity (see Fig.~\ref{fig:tomography01}).

With this measurement, we demonstrate the ability of the ICARUS-Q platform to perform qubit control and measurement. We have also described the implementation of a single qubit algorithm on the RFSoC in the Appendix.

\subsection{Low Noise DC Biasing (Circuit)}
\label{sec:dcbias}
For driving the bias circuit current, needed to set the idle frequency of the tunable qubits, we have developed a low-noise bipolar current source that can be controlled via software from the main computer. 
In the design process we adopted the following considerations. (i) Ultra-low noise: as any noise in the current is directly affecting the coherence properties of the qubit, the current noise of the supply should be as low as possible. (ii) Ultra-low current drift: any drift in the current directly alters the qubit properties and should therefore be suppressed. (iii) Current bandwidth: depending on the design, the change of one flux quantum through the SQUID loop typically corresponds to a change in current of sub-milliampere to a few milliampere. The current range of the source needs to be able to generate a change of at least one flux quantum. In addition the source should be bipolar. (iv) Automation: the source should be addressable via a standard protocol like USB or Ethernet in order to integrate it into the software workflow. 

\begin{figure}[h!]
	\centering
	\includegraphics[width=\linewidth]{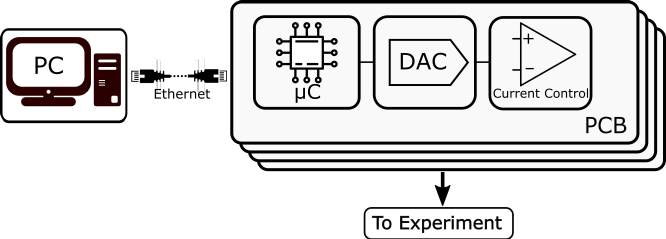}
	\caption{Working principle of the DC current source. A host PC sends current values data to a microcontroller via an Ethernet interface. Subsequently the microcontroller changes the setpoint of the current control circuit via reprogramming the DAC.}
	\label{fig:DCSourceSetup}
\end{figure}

Fig.~\ref{fig:DCSourceSetup} shows the basic design of our current source. The current controller is embedded on a PCB board together with a microcontroller and a DAC that sets the current value. Each PCB hosts 4 current controllers, which are connected to the experiment via SMA cables. In order to set a current value on a certain current controller, the host PC sends the corresponding DAC value and channel number to the microcontroller, which programs the DAC via a serial interface. For the microcontroller we chose a Teensy 4.1 (PJRC.com LLC), which contains an Ethernet interface. Through the Ethernet interface we are able to control multiple PCBs from our host PC.

\subsubsection{Current Controller Circuit}

\begin{figure}[h]
	\centering
	\includegraphics[width=\linewidth]{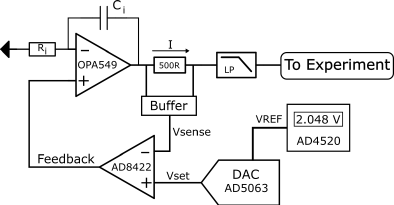}
	\caption{Current control circuit. The output current of an operational amplifier is stabilised by integral feedback. The setpoint is generated by a DAC, which is referenced to a precise voltage reference. Details like supply voltages, decoupling capacitors and connectors have been omitted for clarity.}
	\label{fig:DCSourceCircuit}
\end{figure}

In the following we give a description of the current controller circuit. In the design, we followed the paper by Ref.~\cite{yang2019ultra} with some modifications.
Fig.~\ref{fig:DCSourceCircuit} shows a simplified schematic of the setup.
For clarity, details like supply voltages, decoupling capacitors and connectors have been omitted. 

At the heart of the current controller is an operational amplifier that acts as a current source and is configured as an integrator circuit. We chose to use an OPA547 (Texas Instruments Inc.) for this purpose as it can supply large currents of up to 500~mA and has a low input noise density of $70~\mathrm{nV}/\sqrt{\mathrm{Hz}}$ at 1~kHz. The current output of the amplifier is stabilised using a sense resistor, R$_{\text{sense}}$, that converts the current to a voltage $V_{\mathrm{sense}}$. Subtracting $V_{\mathrm{sense}}$ from a given set voltage, $V_{\mathrm{set}}$, generates a feedback signal that is used for integral control of the current. In this case, the amplifier itself is acting as the integrator. 

It is important that each part in the circuit has low noise and drift. For sensing the current we chose a metal foil resistor (Vishay Foil Resistor) with a resistance of 500 $\Omega$ and a temperature coefficient of $\pm2~\mathrm{ppm}/^\circ$C. To acquire $V_{\mathrm{sense}}$, we first buffer the high-side and low-side of the sense resistor independently using zero-drift operational amplifiers (ADA4522-2, Analog Devices Inc.). The buffered signals are then fed into a precision operational amplifier (AD8422, Analog Devices Inc.) to obtain $V_{\mathrm{sense}}$. For the subsequent determination of the feedback signal another AD8422 chip is used to obtain $V_{\mathrm{sense}}-V_{\mathrm{set}}$.

For generating the set voltage $V_{\mathrm{set}}$, we are using a AD5063 (Analog Devices Inc.) DAC, which has a resolution of 16-bit and is programmable via a serial interface. 
As this current source should be bipolar, we are operating the DAC in its bipolar mode, which allows us to set the output voltage from $-V_{\mathrm{ref}}$ to $+V_{\mathrm{ref}}$, where $V_{\mathrm{ref}}$ is the reference voltage supplied to the DAC. 
It is crucial that $V_{\mathrm{ref}}$ has low noise and low drift, as it directly influences the output current of the source. 
To generate $V_{\mathrm{ref}}$ we are using a ADR4520 (Analog Devices Inc.) voltage reference, which has a temperature coefficient of $\pm2 ~\mathrm{ppm}/^\circ$C and a peak-to-peak noise of 1 $\mu$V in the frequency range of 0.1--10 Hz. 
The output voltage of the ADR4520 chip is $2.048$ V, which results in an output range of $V_{\mathrm{set}}$ from $-2.048$ V to $+2.048$ V.

In our supply design, the current range is given by the sense resistor and the maximum absolute value of $V_{\mathrm{set}}$. For our choice we get a current range of $\pm V_{\mathrm{ref}}/R_{\mathrm{sense}} = \pm 4.096$ mA, which is sufficient for our experiments. We are supplying the current controlling amplifier with a voltage of $\pm 12$ V. This results in a compliance voltage of the supply of $12~\mathrm{V} - V_{\mathrm{ref}} \approx 9.9~\mathrm{V}$. 

\subsubsection{Current Supply Performance}
To demonstrate the noise and stability of the current source, we performed measurements of the current amplitude noise and probed the long-term behaviour by measuring the Allan deviation of the current source. Moreover, we tested the current source in experiment by tuning the frequency of a superconducting qubit. 

\begin{figure}[h]
	\centering
	\includegraphics[width=\linewidth]{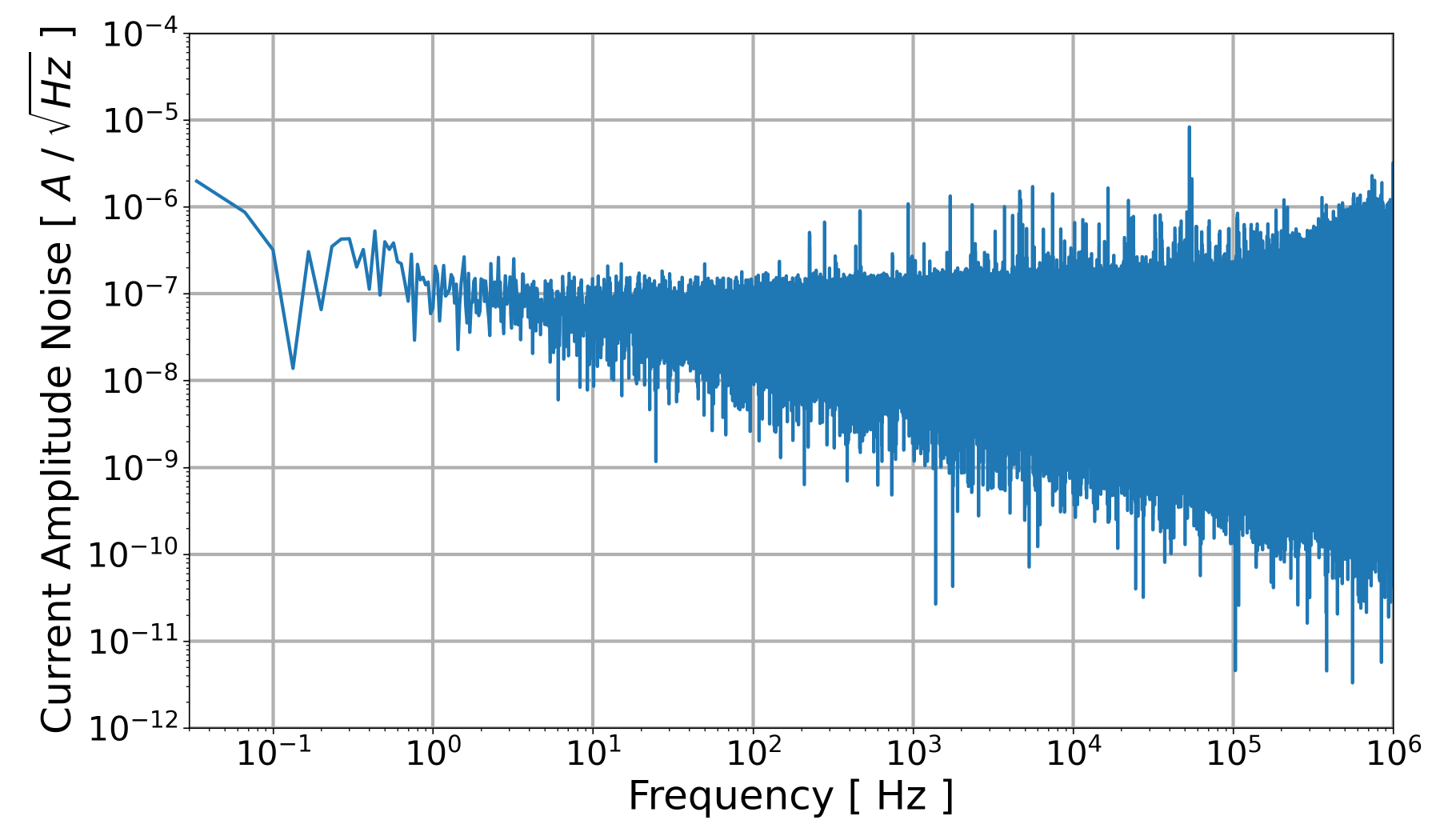}
	\caption{High frequency amplitude noise of the DC current source. The total  noise in the band of 0.1 Hz to 10 kHz is $1.3\times 10^{-7}~\text{A}_{\text{rms}}$. }
	\label{fig:DCSourceNoise}
\end{figure}

We evaluated the current noise and Allan deviation by DC coupling 1~mA of current to the $50~\Omega$ input of a 12-bit digital oscilloscope (LeCroy, Waverunner 610Zi). To determine the current noise, the measured voltage was digitised with a sampling rate of 2~MS/s and a total measurement time of 30~s. The full amplitude range of the oscilloscope was set to 80~mV, which corresponded to a current resolution of 390~nA. From the time series data we calculated the current amplitude noise using Fast Fourier Transform (FFT). Fig.~\ref{fig:DCSourceNoise} shows a typical current amplitude noise spectrum. It can be seen that the broadband noise of the supply is below $3\times 10^{-6}~\mathrm{A}/\sqrt{\mathrm{Hz}}$. Taking the mean value of the amplitude noise in the band of 0.1~Hz--10~kHz and multiplying it by the square root of the same range, we obtain a total noise value of $5.45\times 10^{-6}~\text{A}_{\text{rms}}$ (0.1~Hz--10~kHz).   
Comparing Fig.~\ref{fig:DCSourceNoise} to Ref.~\cite{yang2019ultra}, our setup is most likely limited by the noise floor of the  oscilloscope. This, however,  represents the upper limit of our current source noise density.

\begin{figure}[h]
	\centering
	\includegraphics[width=\linewidth]{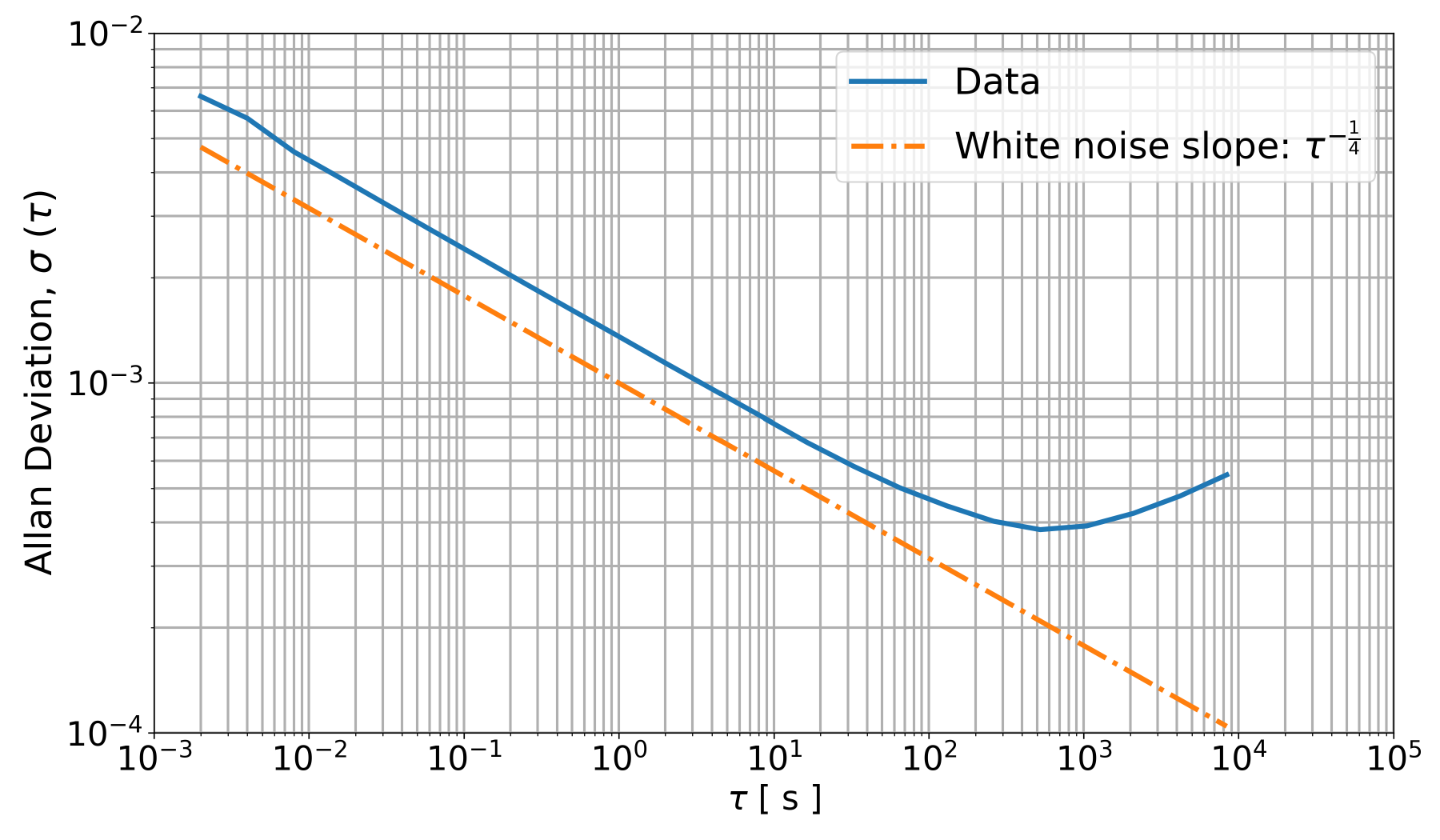}
	\caption{Stability measurement of the current source. Depicted (solid line) is the fractional overlapping Allan deviation of a 18 hours long-term current measurement. The bias stability is about $4\times 10^{-4}$ at 500~s averaging time. Comparison to a white noise slope (dotted-dashed line) shows that for shorter averaging time the deviation is caused by white noise.}
	\label{fig:DCSourceAllan}
\end{figure}

In order to evaluate the stability of the current source, we performed a long-term measurement of the current, using a sampling rate of 500~S/s. The full amplitude range of the oscilloscope was set to 40~mV, which corresponded to a current resolution of 195~nA. From the time series data we calculated the fractional overlapping Allan deviation. Fig.~\ref{fig:DCSourceAllan} (blue line) shows the Allan deviation of the current for a 18 hours long measurement. It can be seen that the bias stability of the source is about $4\times 10^{-4}$ at an averaging time of 500~s. For comparison we also plot the slope of a white noise source (dotted-dashed line). For smaller averaging times, the source of the Allan deviation of the current is consistent with white noise.  

\begin{figure}[h]
	\centering
	\includegraphics[width=\linewidth]{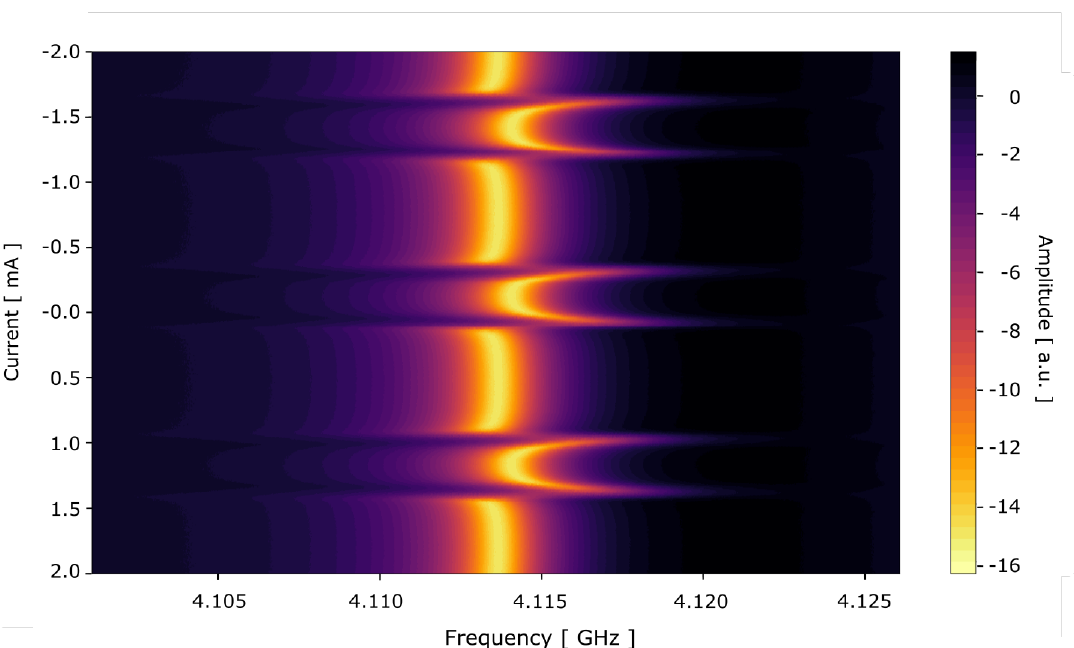}
	\caption{Transmission spectrum of a cavity dispersively coupled to a tunable superconducting transmon qubit. Shown are the spectra for current values from -2.0~mA to 2.0~mA. In this range the frequency of the qubit is going through 3 periodic changes. A continuous wave using a vector network analyzer was used for the readout drive.}
	\label{fig:DCSourceLinearity}
\end{figure}

After characterising the noise performance of the current source, we tested its capabilities of tuning a superconducting circuit. In this case we measured the transmission of a cavity, which was dispersively coupled to a tunable superconducting transmon qubit. Due to the dispersive coupling, a shift of the qubit frequency directly reflects in the shift of the cavity resonance frequency.  
We measured the transmission spectrum of the cavity for various bias currents applied to the qubit. Fig.~\ref{fig:DCSourceLinearity} shows a 2D map of the cavity transmission over a current range of -2.0~mA to +2.0~mA. It can be seen that the frequency of the resonator shifts with applied bias current as expected~\cite{koch2007charge}. Over the measured current range we observe roughly 3 periods of qubit frequency oscillation. 
Measurement of the qubit decoherence as a function of bias current is outside the scope of this work.

\subsection{Software Control \& Cloud Access}
\label{sec:cloud}

In this section we provide an overview of the layout implemented for software
control of the RFSoC systems and the cloud-based execution scheduler of pulse
experiments.

\subsubsection{Software Control}

Following the layout presented in Section~\ref{sec:implementation}, the RFSoC board used in this project has commands for board configuration, external clock locking, channel synchronisation and starting the FPGA.

These commands are executed by a central computer (worker) through the use of the Secure Shell (SSH) protocol. We define a Python class \texttt{IcarusQ} that handles the transfer of user-defined waveforms into the RFSoC, processing of ADC data from the device and execution of the board commands. This class is the interface between experimenter and the RFSoC and provides remote control of the instrument. As this class is written in Python, it can be used alongside other Python instrumentation packages such as \texttt{PyVISA}~\cite{pyvisa},
\texttt{QCoDeS}~\cite{qcodes} and \texttt{Python IVI}~\cite{pythonivi}.

The worker communicates with other instruments such as the trigger
source for the RFSoc boards, via respective connections. Through these connections, the worker is able to control the triggering for the DACs and the ADCs.  On the worker we define a Python function \texttt{IcarusQ-Executor} that runs a pulse experiment. 
This function takes in user-defined trigger timings, number of repetitions and pulse sequence and sets up the trigger control and the RFSoC. Then, it starts both devices and runs the pulse sequence. On completion, it returns the FPGA ADC data as the results of the experiment.

\subsubsection{Cloud Access}

\begin{figure}[h]
	\centering
	\includegraphics[width=\linewidth]{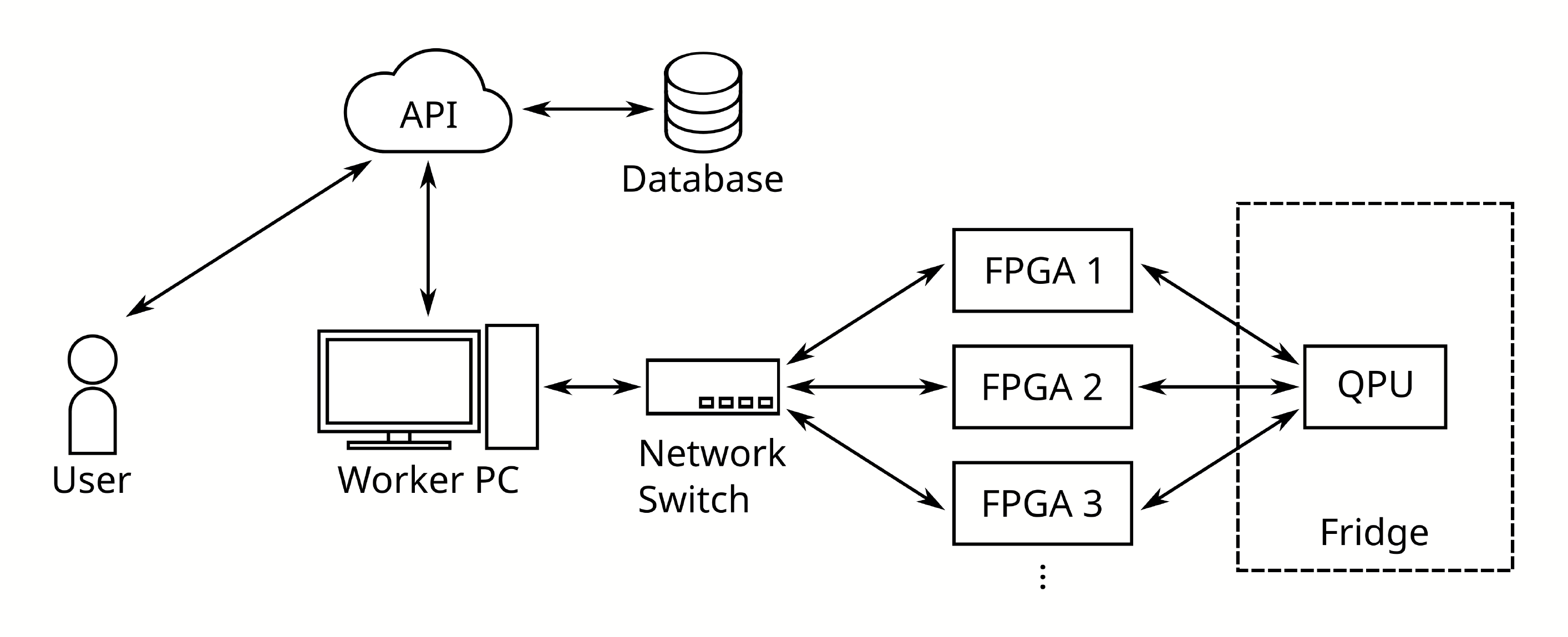}
	\caption{A schematic overview of the cloud access layout
          deployed in this work. Our implementation of cloud-based
          experimental control. A user submits the trigger timings,
          number of repetitions and the pulse sequence to a
          Flask~\cite{flask} server hosted on an Amazon Web Services
          (AWS) EC2 client~\cite{aws}. We provide an Application
          Programming Interface (API) on the Flask server for the user
          to submit and retrieve job results. The job is then stored
          on a Redis~\cite{redis} database. On the worker, we have a
          Python program \texttt{IcarusQ-Scheduler} that queries the
          database for new jobs and runs them with
          \texttt{IcarusQ-Executor}. The results are then uploaded
          onto the Redis database for the user to retrieve.}
	\label{fig:cloud-schematic}
\end{figure}

As the input and output of \texttt{IcarusQ-Executor} is well defined,
we can expose it as a cloud service to run experiments remotely. In
Fig.~\ref{fig:cloud-schematic}, we explain our implementation of
hosting the remote experiment execution on a cloud platform. With this
approach, we have the Flask server~\cite{flask} and Redis~\cite{redis}
database act as a broker between the user and the worker. Neither
party directly communicates with each other and the API follows a
strict format. Hence, we are able to create a secure environment to
execute our experiments remotely.

Finally, the currently layout will be interfaced to the {\tt Qibo}
framework~\cite{efthymiou2020qibo} in order to automate the process
of circuit execution.

\section{Conclusion}
Our setup for superconducting qubit control and measurement was designed with the following goals in mind: to be scalable, to minimise the number of microwave components/instruments to be managed and calibrated, and to have remote access capabilities for the experimenter. 
To these ends, we opted for the Xilinx RFSoC device with sixteen channels of DAC and ADC with high sampling rate to support direct generation and detection of microwave signals at 5--8~GHz. The approach avoids using physical IQ mixers and eliminates the periodic calibration associated with the mixers.  
We also designed a circuit that synchronised several RFSoC boards together using a master clock and triggers for the DAC and ADC channels.
A low-noise DC source was also developed to tune the qubit based on Ref.~\cite{yang2019ultra}.
The remote access to the experiment was implemented with an API to a database on a server, where the user can submit abstracted experimental parameters and retrieve the results.

In the near-term future, a few enhancements to the capabilities are being considered.  
In particular, the use of more features on the RFSoC device such as the utilising the on-board mixer as an alternative waveform generation and sampling method. 
At the moment, the readout signals are processed by the master PC, which can be an issue when dealing with a significantly larger number of qubits. To distribute the data processing load, a future consideration is to use the DSP slices on every RFSoC to process its respective ADC data.
By combining the feedback control feature and the DSP slices as a fast readout system, it may be possible to signal the switching trigger to activate and the correction pulses can be sent to the appropriate channels according to the qubit state~\cite{PhysRevApplied.9.034011}.
The current setup has a relatively low duty cycle for the experiments, which is mostly due to the output data file generation and transferring it out of the RFSoC to the master PC. Improving the experiment duty cycle is currently a work in progress.
Additionally, to further support larger number of samples, we will work on memory optimisation and will also consider faster DDR RAMs (such as QUAD channel RAMs) for direct streaming to/from the channels. 

With all of the important features integrated into a single chip, one can consider the idea of migrating the RFSoC device from room temperature into the dilution refrigerator~\cite{pauka2021cryogenic} for improved SNR and fewer connections to room temperature electronics.
This would provide an alternative implementation of cryogenic control electronics based on commercially available devices, in contrast to other possible technologies such as Josephson arbitrary waveform synthesizer~\cite{9298948} or single-flux technology (SFQ) pulse drivers~\cite{PhysRevApplied.11.014009}.
Next, further developments are to take place in the next generations of RFSoC upon their availability. 

\begin{acknowledgments}
This work was supported by the National Research Foundation Singapore, the Ministry of Education  Singapore, Defence Science and Technology Agency Singapore, DSO National Laboratories Singapore and in part by the Air Force Research Laboratory (AFRL) under contract \#FA9550-19-S-0003.
\end{acknowledgments}

\section*{Author Declarations}
\subsection*{Conflict of Interest}
The authors have no conflicts to disclose.

\subsection*{Author Contributions}
\textbf{Kun Hee Park:}~Conceptualization (equal); Data curation (equal); Formal analysis (equal); Investigation (equal); Methodology (equal); Software (equal); Visualization (equal); Writing - original draft (lead); Writing - review \& editing (lead).
\textbf{Yung Szen Yap:}~Conceptualization (equal); Data curation (equal); Formal analysis (equal); Investigation (equal); Methodology (equal); Software (equal); Visualization (equal); Writing - original draft (lead); Writing - review \& editing (lead).
\textbf{Yuanzheng Paul Tan:}~Conceptualization (equal); Data curation (equal); Formal analysis (equal); Investigation (equal); Methodology (equal); Software (equal); Visualization (equal); Writing - original draft (equal); Writing - review \& editing (equal).
\textbf{Christoph Hufnagel:}~Conceptualization (equal); Formal analysis (equal); Investigation (equal); Methodology (equal); Software (equal); Visualization (equal); Writing - original draft (equal); Writing - review \& editing (equal).
\textbf{Long Hoang Nguyen:} Resources (equal); Software (equal); Visualization (equal).
\textbf{Karn Hwa Lau:}~Software (equal); Writing - review \& editing (equal).
\textbf{Patrick Bore:} Data curation (equal); Formal analysis (equal); Investigation (equal); Software (equal); Writing - review \& editing (equal).
\textbf{Stavros Efthymiou:}~Writing - original draft (equal).
\textbf{Stefano Carrazza:}~Writing - original draft (equal).
\textbf{Rangga P. Budoyo:}~Resources (equal).
\textbf{Rainer Dumke:}~Conceptualization (lead); Funding acquisition (lead); Methodology (lead); Project administration (lead); Supervision (lead); Writing - original draft (equal); Writing - review \& editing (equal).

\section*{Data Availability}
The data that support the findings of this study are available from the corresponding author upon reasonable request.

\appendix

\section*{Appendix}

\subsection*{Gauss Sum Factorization}

Gauss sum factorization is an algorithm using qubit superposition to factorize large numbers~\cite{gauss_cold_atoms}. While the current implementation for superconducting qubits does not yield an advantage over classical factorization algorithms, this algorithm can be used to demonstrate qubit control. 

\begin{figure*}[h!]
    \centering
    \includegraphics[width=\textwidth]{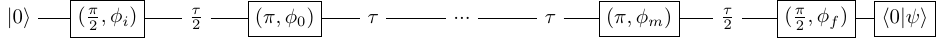}
    \caption{Pulse sequence for Gauss sum factorization. A $\pi/2$-pulse phase shifted by $\pi/2$ is used to rotate the qubit to the superposition state. Next, a train of $m$ $\pi$ pulses is applied to the qubit. The phase of the first $\pi$-pulse is 0 and the $k$-th pulse is $(-1)^k\pi(2k-1)N/l$ where $N$ is the number to factorize and $l$ is the trial factor. Finally, a $\pi/2$-pulse phase shifted by $\pi/2$ is used to rotate the system back to the computational basis. If $l$ is a factor of $N$, the phase will be an integer number of $\pi$ and the $\pi$-pulse train will be in phase, causing the final $\pi/2$ rotation to rotate the system to $\ket{1}$. Otherwise, an arbitrary rotation of the system will occur. The sequence is repeated for $m$ in 1 to $M$, where $M$ is the total number of pulses to be used and the results are averaged by $M$. This is to reduce the impact of arbitrary non-factor rotations that may end up near $\ket{1}$. A delay $\tau$ is applied between pulses.}
    \label{fig:fig1}
\end{figure*}

Using the $\pi$-pulse and $\pi/2$-pulse from the earlier sections, we apply the factorization algorithm (Fig.~\ref{fig:fig1}) using the ICARUS-Q RFSoC for $N=263193$ and $M=15$.

\begin{figure*}[h!]
    \centering
    \includegraphics[width=\textwidth]{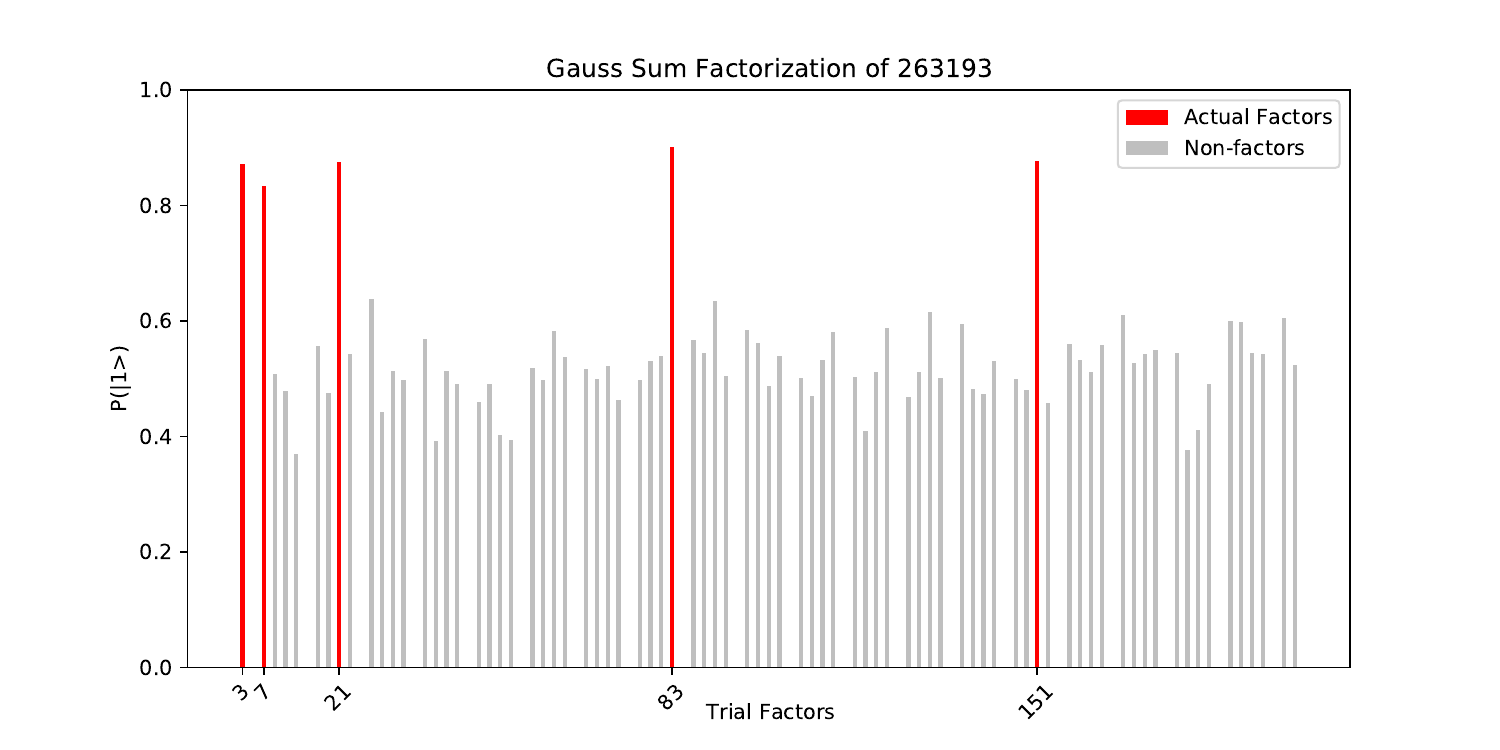}
    \caption{Excited state probability for each trial factors up to 200 in the factorization of $263193$. We apply the preprocessing technique in ~\cite{gauss_ghost_factors} to skip over multiples of $2$ and $5$ as they provide bad contrast compared to the actual factors. The expected factors in this regime are $3, 7, 21, 83, 151$ (highlighted in red). We can see that the major peaks near 1 correspond to the factors, allowing us to identify factors and composite factors. The lowest excited state probability for the factors is 0.834 while the highest excited state probability of the non-factors is 0.639 at $l=27$. The behaviour of this non-factor is discussed in more detail in Ref.~\cite{gauss_ghost_factors}.}
    \label{fig:fig2}
\end{figure*}

We inspect the discernability $\mathcal{D}$ between factors and non-factors. The greater the value of $\mathcal{D}$, the easier it is to identify actual factors over non-factors. $\mathcal{D}$ is determined by the difference between the probabilities for the worst performing factor and the non-factor closest to 1. In an earlier work~\cite{gauss_ghost_factors}, we have calculated the upper limit of $\mathcal{D}$ as $0.67$. In this experiment, we have obtained a value of $\mathcal{D}$ as 0.195 (see Fig.~\ref{fig:fig2}) compared to the previously reported experimental value of 0.4. The difference between these two experiments are the qubits and the number of measurements per data point, which causes the different discernibility. Hence, we have successfully demonstrated the execution of the factorization algorithm on the ICARUS-Q platform.

\clearpage

\bibliography{main}

\begin{thebibliography}{56}
\bibitem [{sec()}]{secondgen}%
  \BibitemOpen
  \href@noop {} {}\bibinfo {note} {{T}he main difference between the first and
  second generation is an extra 1~{GH}z in the maximum input
  frequency.}\BibitemShut {Stop}%
\bibitem [{mem()}]{memoryprob}%
  \BibitemOpen
  \href@noop {} {}\bibinfo {note} {In the product description, there is a total
  of 60.5 Mbits memory. This is actually a combination of 38.0 Mb of block RAM
  and 22.5 Mb of UltraRAMs. In our current design, we are only able to utilize
  the block RAM for various purposes and not solely for the ADC and DAC
  data.}\BibitemShut {Stop}%
\end{thebibliography}

\end{document}